\documentclass[twocolumn,amsmath,amssymb,osajnl]{revtex4}
\usepackage{epsfig}
\usepackage{verbatim}

\newcommand{\EEE}{\mathbf{E}}
\newcommand{\DDD}{\mathbf{D}}
\newcommand{\rrr}{\mathbf{r}}

\newcommand{\ud}{{\rm d}}

\begin{document}

\title{Sensitivity of photonic crystal fiber grating sensors: biosensing, refractive index,  strain, and temperature sensing}

\author{Lars Rindorf and Ole Bang}

\affiliation{COM$\bullet$DTU, Department of Communications,
Optics and Materials, Technical University of Denmark, DK-2800 Kgs. Lyngby, Denmark}

\begin{abstract}
We study the sensitivity of fiber grating sensors in the
applications of strain, temperature, internal label-free biosensing,
and internal refractive index sensing. It is shown that optical
dispersion plays a central role in determining the sensitivity, and
the dispersion may enhance or suppress sensitivity as well as change
the sign of the resonant wavelength shifts. We propose a quality
factor, $Q$, for characterizing LPGs.
\end{abstract}


\maketitle

\section{Introduction}
Optical fiber grating sensors are becoming increasingly widespread
in a wide range of applications \cite{lee2003}. Their
characteristics include immunity to electromagnetic interference,
high sensitivity, fast response, and keeping their calibration with
time. Furthermore they possess a high tolerance to harsh
environments such as sea water, concrete, and extreme temperatures.
The measured parameter, the resonant wavelength of the grating, is
usually linear in the magnitude of the measurand, e.g. temperature
or strain, allowing for an accurate quantitative estimate of the
measurand.

Photonic crystal fibers (PCFs) \cite{birks1997} are optical fibers,
in which the cladding consists of a microstructured array of air
holes running along the fiber axis. Here we consider index guiding
PCFs, which have a solid core in the form of a missing air hole in
the center. In index guiding PCFs light is confined to the core by
modified total internal reflection, analogous to guiding in standard
optical fibers. The holes in the cladding are most commonly placed
in a periodic triangular structure, which is characterized by a
lattice constant (pitch) and the diameter of the holes. PCFs are
typically made of silica \cite{birks1997} or, more recently, polymer
materials, where they are termed microstructured polymer optical
fibers (mPOFs) \cite{eij2001}. The hole structure in the cladding
determines the optical properties of the PCF, which allows for a
large degree of freedom in tayloring key properties, such as
dispersion and group velocity.

In standard optical fibers the silica in the core contains dopants
to increase the refractive index. The refractive index of the doped
silica is sensitive to UV radiation and this may be used for
inscription of gratings. This technique can not be used for pure
silica PCFs, since the core is not UV sensitive. For long-period
gratings (LPGs) in PCFs other techniques exist, such as using an
electric arc \cite{humbert2003}, a CO$_2$ laser
\cite{kakarantzas2002}, or mechanical microdeformation
\cite{nielsen2003}. Otherwise the PCF can be specially made with a
UV sensitive core for fiber gratings allowing for the inscription of
both LPGs and Bragg gratings (BGs) \cite{eggleton:1999}.

Polymer has a smaller Young's modulus and greater elastic limit than
silica, which makes it a good material for measuring strain and
bending. In addition, polymer is biocompatible and can bind
biomolecules directly to its surface, making it ideal for
biosensing. Standard polymer optical fibers are made of poly-methyl
methacrylate (PMMA), which is inherently photosensitive, allowing
for both BGs \cite{peng1999} and LPGs \cite{li2005} to be inscribed
using a UV lamp. The inherent UV sensitivity was also recently used
to write BGs in mPOF \cite{dobb2005}, whereas heat treatment under
mechanical stress was used to write the first permanent LPG in an
mPOF \cite{hiscocks2006}.

An important property of grating sensors in silica PCFs are that
they are largely temperature insensitive, showing $\sim$ 1 pm/K for
both BGs \cite{frazao2005} and LPGs \cite{dobb2004}. The strain
sensitivity is of the same order for LPGs and BGs in PCFs, and thus
strain sensors may be realized with reduced cross sensitivity to
temperature sensing. In addition, the holes of a PCF can be
infiltrated with a substance having a high temperature coefficient
to reduce or enhance the temperature sensitivity
\cite{sorensen2006}.

Whereas fiber-optic strain and temperature sensors use the material
properties of the fiber, fiber-optic biosensors and refractive index
sensors use the principle of evanescent wave sensing. In particular,
there is a growing interest in optical biochemical sensors
\cite{delisa2000}. Here PCFs have the advantage that the analyte may
be infiltrated into the air holes and thereby create a strong
interaction with the probing electromagnetic field
\cite{jensen2004a,jensen2005}. When using mPOFs one can take
advantage of different polymers with special biosensing properties,
such as TOPAS \cite{gem2007}.

 Liquid infiltrated PCF-LPGs \cite{kerbage2003,rindorf2006_7} and PCF-BGs
\cite{phanhuy2006,sorensen2006} have been demonstrated, which is an
important step towards the ultimate goal of developing a so-called
\emph{label-free} biosensor in a PCF. Fiber gratings potentially
allow for label-free biosensing with PCFs \cite{rindorf2006_7} by
tracking how the grating resonance changes during operation. This
may open for new possibilities in biomedical applications
\cite{rindorf2006_6}.

The dispersion plays a crucial role for all types of grating sensors
in standard optical fibers \cite{acharya,ramachandran2005}. The
dispersion of PCFs depends strongly on the hole structure in the
cladding and thus the performance of the PCF grating sensor can be
designed by an appropriate choice of pitch and hole size.

In this paper we explore the possibility of optimizing the hole
structure of PCFs for sensing purposes. We do this by analyzing the
sensitivity of triangular PCF gratings for different pitch and hole
diameters. The applications under study are biosensing, refractive
index sensing, strain, and temperature sensing. We show that the
sensitivity is strongly influenced by the choice of the structural
parameters of the PCF. We employ realistic values for the fibers
parameters and give realistic values of pitch and hole diameter that
minimize or maximize the sensitivity to different measurands.

We find that BGs are best characterized in the usual way by their
sensitivity, $\gamma$, defined as the resonant wavelength shift
divided by the resonant wavelength. In contrast, we find that the
LPGs are best characterized by their quality factor, $Q$, defined as
the resonant wavelength shift divided by the full-width at half
maximum (FWHM) of the resonance dip.
\section{Photonic crystal fibers}
\begin{figure}[b!]
\begin{center}
\includegraphics[angle = 0, width =0.45\textwidth]{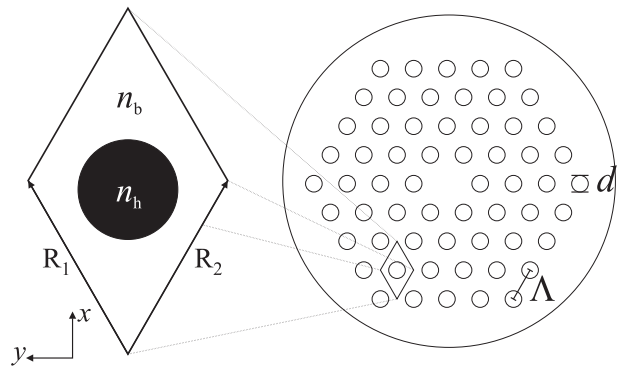}
\end{center}
\caption{A PCF with triangular cladding structure. The structure is
characterized by the lattice vectors, $R_1$ and $R_2$, the hole
size, $d$, and the pitch, $\Lambda$. The refractive  indices $n_h$
and $n_b$ are the hole and base material refractive
indices}\label{fig:analyte}
\end{figure}

The eigenmodes of the PCFs for a temporally harmonic electrical
field, $\EEE(\rrr,t)=\EEE_\omega(\rrr)e^{-i\omega t}$, may be found
from Helmholtz eigenvalue equation,
\begin{eqnarray}
\nabla\times\nabla\times \EEE_\omega(\rrr_\bot,z) =
\frac{\omega^2}{\textrm{c}^2}\varepsilon(\rrr_\bot)\EEE_\omega(\rrr_\bot,
z) \label{eq:eigenvalue},
\end{eqnarray}
where $\omega$ is the angular frequency, c is the speed of light in
vacuum, $\varepsilon$ is the dielectric function, and
$\rrr_\bot=(x,y)$ is the transverse coordinate vector. The $z$-axis
is chosen to coincide with the fiber axis. We look for plane-wave
solutions of the form $\EEE_\omega(\rrr) =
\EEE_\omega(\rrr_\bot)e^{i\beta z}$ with $\beta$ being the
propagation constant. The effective index is defined as
$n_\textrm{eff}(k) = \beta(k)/k$, where $k$ is the free-space
wavevector and is related to the frequency and wavelength by $k =
\omega/\textrm{c} = 2\pi/\lambda$. The effective indices of the core
mode and a resonant cladding mode are denoted by $n_\textrm{co}$ and
$n_\textrm{cl}$. The orthonormalization
 for Eq.~(\ref{eq:eigenvalue}) is
\begin{eqnarray}
\int_\Omega \ud\rrr_\bot \EEE^\dagger_m(\rrr_\bot)
\varepsilon(\rrr_\bot) \EEE^\dagger_n(\rrr_\bot) = \delta _{mn}.
\end{eqnarray}
 The PCF is typically made of two materials with a high refractive
index contrast. A general dielectric function, $\varepsilon(\rrr)$,
 for a two material triangular PCF
structure is described by high and low index regions as seen in Fig.
\ref{fig:analyte}. The periodic triangular structure is spanned by
the lattice vectors $R_1 = (\sqrt{3}, 1)/2$ and  $R_2 = (\sqrt{3},
-1)/2$.  The general dielectric function equals $n_b^2$ for the base
region and $n_h^2 $ for the hole regions, with $n_b$ and $n_h$ being
the refractive indices of the base and hole material, respectively.

The effective index of the core depends strongly on the wavelength,
the hole diameter, and the pitch. The hole diameter, $d$, and pitch,
$\Lambda$, are indicated in Fig. \ref{fig:analyte}. Large hole
diameters and small pitch also give a small core and therefore a
strong effective nonlinearity.

Triangular silica PCFs are single mode for $d/\Lambda < 0.42$ for
all wavelengths [\onlinecite{kuhlmey2002,mortensen2003c}].  For
small hole sizes ($d/\Lambda < 0.3$) the PCF is only weakly guiding,
and bend losses increase drastically. Bend losses also become
pronounced in the short wavelength limit, $ \lambda \ll \Lambda$,
regardless of hole diameter. PCFs can be designed to have a large
mode area.

The transmission spectrum of an LPG in a PCF only shows a few
cladding mode resonances, whereas several resonances are seen in a
standard optical fiber LPG spectrum. The cladding modes of a PCF are
different from the cladding modes of a standard optical fiber, since
the fiber cross sections obey very different symmetries. In
particular a PCF obeys a six-fold rotational symmetry ($C_6^v$)
while the standard optical fiber has circular symmetry. The
effective indices of the cladding modes of a PCF are closely spaced,
and this spacing is much smaller than the difference between the
core effective index and the effective index of any cladding mode.
 In other words, the beat length between any two cladding modes is
much longer than the beat length between a cladding mode and the
core mode. In practice we can assume $n_{\textrm{cl}}\simeq
n_\textrm{FSM}$ for all cladding modes, where the fundamental space
filling mode, $n_\textrm{FSM}$, is the effective index of an
infinite cladding structure spanned by the lattice vectors in
Fig.~\ref{fig:analyte}.

\begin{figure}[b!]
\begin{center}
\includegraphics[angle = 0, width =0.45\textwidth]{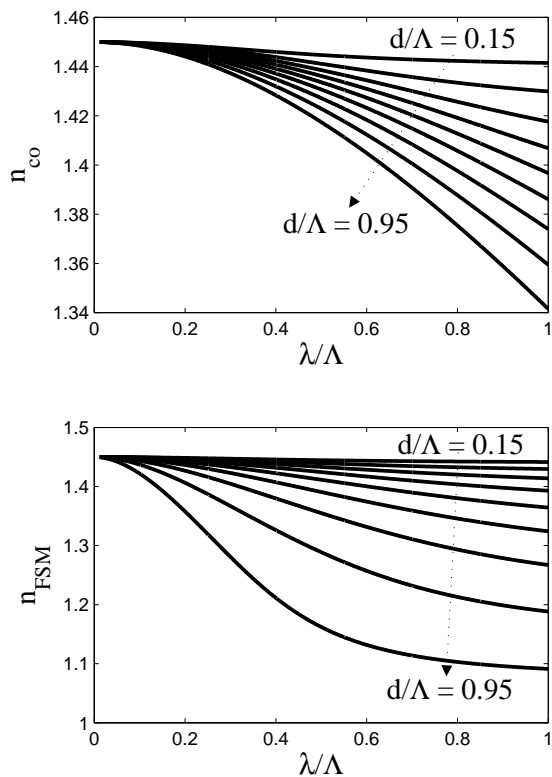}
\end{center}
\caption{Effective indices for the core (top) and fundamental space
filling mode (FSM) modes for triangular photonic crystal fibers. The
lines indicate different values of the hole diameter relative to the
pitch. The lines correspond to different values of $d/\Lambda$,
increasing by 0.1}\label{fig:neffs}
\end{figure}

We have calculated the effective index for the core mode,
$n_\textrm{co}$, and the fundamental space filling mode,
$n_\textrm{FSM}$, for different hole to pitch and wavelength to
pitch ratios, with the base material having a refractive index
$1.45$, which is the refractive index of silica at the wavelength
$\lambda \simeq 1\mu$m. The effective indices are shown in
Fig~\ref{fig:neffs}. The effective indices decrease with increasing
hole size ($d/\Lambda$) or increasing normalized wavelength
($\lambda/\Lambda$). The effective index of the core mode is larger
than that of the FSM, regardless of the fiber structure parameters
($d$,$\Lambda$).

\section{Coupled mode theory}
\subsection{Introduction}
Coupled mode theory  \cite{yariv_book} is a simple and compact
theory that has been successful in accounting for the
characteristics of fiber gratings. It allows us to derive the
qualitative behavior of the gratings with some general, simplifying
assumptions on the physical mechanisms. The theory assumes that near
resonance the electromagnetic field propagating through the fiber
may be expressed as a superposition of eigenmodes of the unperturbed
fiber
%
$a_\textrm{co}(z)\EEE_\textrm{co}(\rrr_\bot)e^{i\beta_\textrm{co}z}
+ a_i(z)\EEE_i(\rrr_\bot)e^{i\beta_iz}
$, 
 where `$i$' denotes a resonant mode, and the $a$'s are amplitude
coefficients that depend on $z$ due to the perturbation. For Bragg
gratings the resonant mode is the counter propagating core mode, and
for long-period gratings the resonant mode is a copropagating higher
order mode, usually a cladding mode.
 The coupled-mode equations constitute a set of linear coupled differential equations. They may be expressed in a compact
  matrix form as
\begin{eqnarray}
\left[ \begin{array}{c} \frac{\partial a_\textrm{co}(z,\delta_i)}{\partial z} \\
\frac{\partial a_i(z,\delta_i)}{\partial z} \end{array} \right] =
-i\left[ \begin{array}{cc} 0 & \kappa_i e^{i\delta_i z} \\
\kappa_i^* e^{-i\delta_i z} & 0 \end{array} \right] \left[
\begin{array}{c} a_\textrm{co}(z,\delta_i)\\ a_i(z,\delta_i) \end{array} \right]
\label{eq:cmt}
\end{eqnarray}
where the detuning in the resonance between the core mode and the
resonant mode, $i$, is
\begin{eqnarray}
\delta_i \equiv \beta_\textrm{co}-\beta_i -
\frac{2\pi}{\Lambda_\textrm{G}} \label{eq:detuning}
\end{eqnarray}
 and the coupling constant is
\begin{eqnarray}
\kappa_i =
\frac{k^2}{2\sqrt{|\beta_\textrm{co}\beta_\textrm{i}|}}\int_\Omega
d\rrr_\bot \EEE^\dagger_\textrm{co}(\rrr_\bot) \Delta
\tilde\varepsilon\big(\rrr_\bot\big)\EEE_i (\rrr_\bot),
 \label{eq:couplingconstant}
\end{eqnarray}
where the integration is over the fiber cross section, $\Omega$.
$\Delta \tilde\varepsilon\big(\rrr_\bot\big)$ is the first phase
matched Fourier component ($\pm 1$) in the  expansion
$\Delta\varepsilon (\rrr_\bot,z) = \sum_{m= -\infty}^\infty
\Delta\tilde \varepsilon_m (\rrr_\bot) \exp( - i
m\frac{2\pi}{\Lambda_\textrm{G}}z )$.

In Eq.~(\ref{eq:cmt}) we have neglected the diagonal terms, the so
called self-coupling terms. The self-coupling arises from the
perturbation of the individual modes due to the index change of the
grating. The self-coupling shifts the resonant wavelength some nm's,
while otherwise leaving the spectrum of the grating unaltered. Since
this effect is typically small we neglect it for our purposes.

\begin{figure}[b!]
\begin{center}
\includegraphics[angle = 0, width =0.5\textwidth]{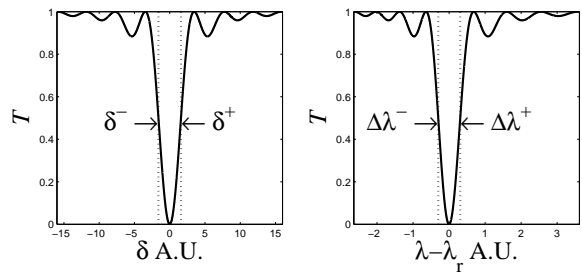}
\end{center}
\caption{Definition of the full-width half maximum (FWHM) in
detuning, $\delta_\textrm{FWHM} = \delta^+-\delta^-$, and in
wavelength, $\lambda_\textrm{FWHM} =
\Delta\lambda^+-\Delta\lambda^-$.
  }\label{fig:fwhm}
\end{figure}

\begin{figure}[b!]
\begin{center}
\includegraphics[angle = 0, width =0.5\textwidth]{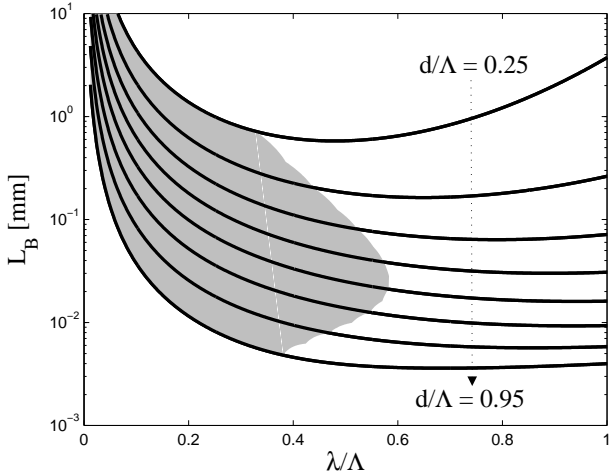}
\end{center}
\caption{ Beat length as function of the wavelength over pitch,
$\lambda/\Lambda$, for an LPG. The gray area indicates negative
group index mismatch $\bar n _\textrm{g} < 0$}\label{fig:lg}
\end{figure}

\begin{figure}[b!]
\begin{center}
\includegraphics[angle = 0, width =0.5\textwidth]{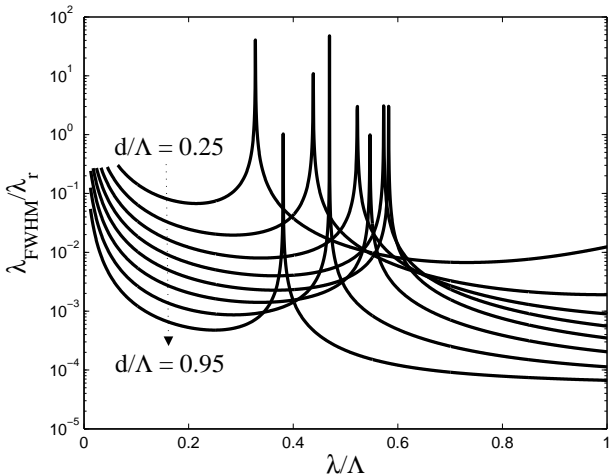}
\end{center}
\caption{FWHM as function of the wavelength over pitch,
$\lambda/\Lambda$ for an LPG }\label{fig:fwhm2}
\end{figure}

Using Eq.~(\ref{eq:detuning}) we obtain a resonance condition in
terms of effective indices and the wavelength
\begin{eqnarray}
\lambda_\textrm{r} &=& \big(n_\textrm{co}(\lambda_\textrm{r})
-n_{i}(\lambda_\textrm{r})\big) \Lambda_\textrm{G}\nonumber
\\
&\equiv & \bar n_\textrm{f}  (\lambda_\textrm{r}) \Lambda_\textrm{G}
\label{eq:res_LPG} \label{eq:res_BG} ,
\end{eqnarray}
where $n_i$ is the effective index of the resonant mode. The
equation is valid both for LPGs,  $\bar n_\textrm{f} = n_\textrm{co}
- n_\textrm{cl}$, and BGs, $\bar n_\textrm{f} = 2\,n_\textrm{co}$.

In the following we use the bar for indicating a difference between
the core and the resonant modes as done in Eq.~(\ref{eq:res_LPG}).
Thus $\bar \beta \equiv \beta_\textrm{co}-\beta_i$ for the
propagation constant, and for the group index we have
$\frac{\partial \bar \beta}{\partial k} =
n_\textrm{g,co}-n_{\textrm{g},i}\equiv \bar n_\textrm{g}  =
-\lambda^2 \partial_\lambda (1/L_B)$, where $L_B =2\pi/\bar\beta$ is
the beat length, with
 $n_{\textrm{g},i} \equiv
\frac{\partial \beta_i}{\partial k}$.

The resonance conditions must be solved considering the dispersion
of the effective indices, since these are strongly dependent on the
pitch and hole size of the PCF structure  as seen in
Fig.~\ref{fig:neffs}.

\subsection{Dispersion effects on the wavelength FWHM}
Couple-mode theory in the form of Eq.~(\ref{eq:cmt}) will give an
expression for the transmission in terms of the detuning and
coupling constant. However, experimentally the transmission is
measured as function of wavelength.

In Fig.~\ref{fig:fwhm} the FWHM for the detuning,
$\delta_\text{FWHM} = |\delta^+ - \delta^-|$, and the wavelength,
$\lambda_\text{FWHM} = |\Delta\lambda^+ - \Delta\lambda^-|$, are
defined. By definition we have $T(\delta^+) = T(\delta^-) =
0.5$, and equivalently for the wavelength we have
$T(\lambda_\textrm{r} + \Delta\lambda^+) =T(\lambda_\textrm{r} +
\Delta\lambda^-) = 0.5$.

 To find a relation between $\delta_\textrm{FWHM}$ and $\lambda_\textrm{FWHM}$ that
incorporates the dispersion of the effective indices we expand the
detuning around the resonance wavelength to first order in
$\Delta\lambda$
\begin{eqnarray}
\delta(\lambda_\textrm{r}+\Delta\lambda) &=&
\beta_\textrm{co}(\lambda_\textrm{r}+\Delta\lambda)-\beta_{i}(\lambda_\textrm{r}+\Delta\lambda)
- \frac{2\pi}{\Lambda_\textrm{G}} \nonumber \\ &\simeq&
 \frac{\partial \bar \beta}{\partial k}\frac{\partial k}{\partial \lambda}\Big|_{\lambda_\textrm{r}}\Delta\lambda
  = -\bar
n_\textrm{g}(\lambda_\textrm{r})\frac{2\pi}{\lambda_\textrm{r}^2}\Delta\lambda,
 \label{eq:deltalambda}
\end{eqnarray}
With this equation it is possible to obtain two equations relating
$\delta^+$ and $\delta^-$ to $\Delta\lambda^+$ and
$\Delta\lambda^-$. Subtracting the two equations and taking the
absolute value of both sides, we obtain an expression for
$\delta_\text{FWHM}$ and $\lambda_\text{FWHM}$
\begin{eqnarray}
\frac{\lambda_\textrm{FWHM}} {\lambda_\textrm{r}} &=&
\frac{1}{2\pi}\frac{\lambda_\textrm{r}\delta_\textrm{FWHM} }{|\bar
n_\textrm{g}(\lambda_r) |}  \label{eq:fwhm},
\end{eqnarray}
which incorporates the dispersion of the effective indices in a
consistent manner.

We note that the expression is not valid at group index matching
between the two resonant modes, $\bar n_\textrm{g}=0$. In this
limit, however, the FWHM goes to infinity, and any wavelength shifts
of the sensor will become difficult to detect.

\subsection{Long-period gratings}\label{sec:FWHM_LPG}
Long-period gratings couple the core mode to a co-propagating
high-order mode. The difference in effective index of two such modes
is typically small, hence by Eq.~(\ref{eq:res_LPG}) the grating
period of such a grating must be long, $\Lambda_\textrm{G} \sim
100-1000 \times \lambda$. For LPGs we put the subscript $i$ = FSM,
and $a_i = a_\textrm{FSM}, \delta = \delta_i = \delta_\textrm{FSM},
\kappa = \kappa_i = \kappa_\textrm{FSM}$

 Typically the FWHM of an
LPG is of the order of nanometers. The minimal detectable shift in
resonance condition is thus set by the FWHM and not the spectral
resolution of the spectrum analyzer.

The coupled-mode equations for the long-period grating may be solved
analytically from Eq.~(\ref{eq:cmt}) yielding the transmission
coefficient
\begin{eqnarray}
T(\delta) = 1- \frac{1}{1+\frac{\delta^2}{4|\kappa|^2}
}\sin^2\Big(|\kappa|L\sqrt{1+ \frac{\delta^2}{4|\kappa|^2}}\Big).
\end{eqnarray}
At resonance ($\delta = 0$) the transmission is sinusoidal, $
T(\delta = 0) = \cos^2(|\kappa|L)$, where $L$ is the length of the
grating. The first zero for $T$ is at $|\kappa|L = \pi/2$. Ignoring
the wavelength dependence of the coupling constant, $\kappa =
\kappa(\lambda_\textrm{r}) = \pi/(2L)$, we substitute $\xi =
\delta/(2|\kappa|)$ and solve $T(\xi) = 0.5$ numerically. This
yields $\xi^+ = -\xi^- \simeq 0.80$. We can find the FWHM of the
detuning as ${\delta_\textrm{FWHM}}/{|\kappa|} = 2(\xi^+ -\xi^-)
\simeq 3.2$. The wavelength FWHM of the
transmission can then be found using Eq.~(\ref{eq:fwhm})
\begin{eqnarray}
\frac{\lambda_\textrm{FWHM}} {\lambda_\textrm{r}} &\simeq& 0.80
\frac {1}{|\bar n_\textrm{g}(\lambda_\textrm{r})|} \frac
{\lambda_\textrm{r}}{L} . \label{eq:lpg_fwhm}
\end{eqnarray}
The wavelength FWHM obviously is large when the group index matching
is close to zero, $\bar n_\textrm{g}(\lambda_\textrm{r}) = 0$.

In conclusion we have found an expression, characteristic for LPGs,
that relates $\lambda_\textrm{FWHM}$ directly with the length of the
grating, the group index and the resonant wavelength. The expression
is derived under the assumption, that $ |\kappa(\lambda)L = \pi/2$

\subsection{Bragg gratings}
Bragg gratings couple the incident core mode to a
counter-propagating core mode, hence $\beta_i = -\beta_\textrm{co}$
and the effective index is $n_{i} = n_{\rm co}$ and the coupling
constant $\kappa = \kappa_{i} = \kappa_{\rm co}$. From
Eq.~(\ref{eq:res_LPG}) we find that the period of such a grating
must be comparable with the wavelength, $\Lambda_\textrm{G} \sim
\lambda/2$. The FWHM of a BG is usually very narrow compared with an
LPG. Although a narrow FWHM gives a high resolution for a sensor, it
also puts high demands on production tolerances and the optical
spectrum analyzer. Thus the minimal detectable shift in the resonant
wavelength is set by the spectrum analyzer.

Fluctuations in the effective index caused by imperfections in the
PCF structure or the grating period along the grating enters by
Eq.~(\ref{eq:res_BG}) in the transmission spectrum and can inflict a
significant broadening of the resonance peak \cite{phanhuy2006}. On
the beneficial side one need only to couple light into one end of
the PCF, since it is possible to measure the reflection spectrum of
the grating.

The coupled-mode equations for the Bragg grating may be solved
analytically from Eq.~(\ref{eq:cmt}) yielding a transmission
coefficient
\begin{eqnarray}
T(\delta) = 1-\frac{\sinh^2\Big(|\kappa| L\sqrt{1 -
\frac{\delta^2}{4|\kappa|^2} }\Big)}{ \cosh^2\Big(|\kappa| L\sqrt{1
- \frac{\delta^2}{4|\kappa|^2}}\Big)- \frac{\delta^2}{4|\kappa|^2}}.
\label{eq:BG_T_uni}
\end{eqnarray}
 At resonance the transmission is equal to
$T(0) = 1- \tanh^2(|\kappa|L) $. Assuming that the coupling constant
is constant, $\kappa = \kappa(\lambda_\textrm{r})$, we find
numerically that the $\delta_\textrm{FWHM}$ can be approximated
reasonably well by
\begin{eqnarray}
\delta_\textrm{FWHM}  \simeq  \frac{4.0|\kappa|L +  3.4}{ L}.
\label{eq:BG_FWHM_L}
\end{eqnarray}
According to this estimate the FWHM decreases with the grating
length until a lower bound set by the coupling strength itself.

Typically $|\kappa|L > 1$ and we can neglect the second term in
Eq.~(\ref{eq:BG_FWHM_L}) and obtain
 $\delta_\textrm{FWHM} \simeq 4|\kappa|$.
  The wavelength FWHM  is then
\begin{eqnarray}
\frac{\lambda_\textrm{FWHM}} {\lambda_\textrm{r}} &\simeq& 0.64
\frac{\lambda_\textrm{r}|\kappa|}{|\bar
n_\textrm{g}(\lambda_\textrm{r})|}. \label{eq:bg_fwhm}
\end{eqnarray}
We note that the FWHM generally increases with the resonance
wavelength squared.

In conclusion we have found an expression, characteristic for BGs,
that relates $\lambda_\textrm{FWHM}$ directly with the coupling
constant, the group index of the core mode and resonant wavelength.
We have made the assumption the coupling constant is independent of
wavelength.

\section{Linear response theory for fiber sensors}
The fiber sensor is perturbed by a measurand to produce a change in
the measured quantity, in our case the resonant wavelength. For
quantitative and qualitative estimation of the  measurand a linear
relation to the resonant wavelength is assumed. Since the resonant
wavelength shifts are generally small compared to the wavelength, it
is natural to employ perturbation theory and Taylor expand the
resonance condition to first order around the resonance wavelength.

Consider a general measurand, $\alpha$. We assume that the shift in
resonant wavelength as well as in the measurand are sufficiently
small, and we expand the resonance condition Eq.~(\ref{eq:res_LPG})
in the resonant wavelength and the measurand to first order
\begin{eqnarray}
\lambda_\textrm{r}+\Delta\lambda_\textrm{r}  = \Big[\bar
n_\textrm{f}(\lambda_\textrm{r},\alpha) + \frac{\partial \bar
n_\textrm{f}}{\partial \lambda_\textrm{r}} \Delta\lambda_\textrm{r}
+ \frac{\partial \bar n_\textrm{f}}{\partial \alpha}
\Delta\alpha\Big]\Lambda_\textrm{G}.
\end{eqnarray}
Identifying $\Lambda_\textrm{G} \equiv \lambda_\textrm{r}/\bar
n_\textrm{f}(\lambda_\textrm{r})$ by Eq.~(\ref{eq:res_BG}) and $\bar
n_\textrm{g} =  \bar n_\textrm{f} - \lambda \frac{\ud \bar
n_\textrm{f}}{\ud\lambda}$, we obtain an expression for the
sensitivity
\begin{eqnarray}
\gamma  &=& \Big|\frac{1}{\lambda_\textrm{r}}\frac{\ud
\lambda_\textrm{r}}{\ud \alpha}\Big| = \Big|\frac {1} {\bar
n_\textrm{g}(\lambda_\textrm{r})}\frac{\partial \bar
n_\textrm{f}}{\partial \alpha}\Big|\nonumber  \\ &=& \left\{
\begin{array}{cl}
\frac{1}{|n_\textrm{g,co} - n_\textrm{g,cl}|}\Big|\frac{\partial (n_\textrm{co} - n_\textrm{cl})}{\partial \alpha}\Big| , & \qquad \textrm{LPG},\\
\frac{1}{|n_\textrm{g,co}|}\Big|\frac{\partial
n_\textrm{co}}{\partial \alpha}\Big|, & \qquad \textrm{BG}.
\label{eq:gamma}
\end{array}\right.
\end{eqnarray}
The shift in the resonant wavelength is thus dependent on the change
in the difference of the effective indices relative to the
difference in group indices. We have chosen the symbol `$\gamma$'
for the sensitivity to match the notation of Shu et
al.~\cite{shu2002}. Obviously, when $|\gamma|$ is large the sensor
will give large wavelength shifts, which will give a high
sensitivity and accuracy of the measurand. However, changing the
fiber parameters not only changes the wavelength shifts but also
change the full-width half maximum (FWHM) of the resonance dip.
Thus, even though $|\gamma|$ may be large for a certain fiber
structure, the wavelength shifts will be difficult to detect if they
are much smaller than the FWHM of the resonance dip. In particular,
we therefore note that the sensitivity is high close to group index
matching, $\bar n _\textrm{g} = 0$, which by
Eqs.~(\ref{eq:lpg_fwhm},\ref{eq:bg_fwhm}) is also where the FWHM is
wide. We introduce the quality factor, $Q$, of the grating sensor,
which is the wavelength shift divided by the FWHM,
\begin{eqnarray}
Q &=& \frac{1}{\lambda_\textrm{FWHM}}\frac{\ud
\lambda_\textrm{r}}{\ud \alpha}
\nonumber\\
 &\simeq&
\left\{
\begin{array}{cl}
 \frac {1} {0.80 } \frac{L}{\lambda_\textrm{r}}\frac{\partial (n_\textrm{co} - n_\textrm{cl})}{\partial \alpha}, &\qquad \textrm{LPG},\\
 \frac{1}{0.64  } \frac{1}{\lambda_\textrm{r}|\kappa|}\frac{\partial n_\textrm{co}}{\partial \alpha}, &\qquad
 \textrm{BG}.
\end{array}\right.\label{eq:q}
\end{eqnarray}
 We note
that $Q$, in contrast to $\gamma$, does not increase when $ \bar
n_\textrm{g} \to 0$.
 Perfect group index matching $ \bar n_\textrm{g} = 0$ only occurs for LPGs, as we will see in the
following.

We have derived two expressions for the performance of a fiber
grating sensor. The sensitivity, $\gamma$, is inversely proportional
to the group index mismatch, $\bar n_{\textrm g}$, and expresses the
shift in the resonant wavelength relative to the resonant
wavelength. The quality factor, $Q$, in contrast, expresses the
wavelength shift relative to the FWHM. Ideally both the sensitivity,
$\gamma$, \emph{and} the quality factor, $Q$, should be large.

The FWHM of LPGs is usually larger than the resolution of the
spectrum analyzer.  The quality factor is thus the most important
 measure of the performance of LPGs. The FWHM of BGs is usually narrow, and comparable to the
resolution of the spectrum analyzer. The sensitivity is then the
most important for BGs.

\section{Numerical simulations}
\subsection{Effective indices}
We employ a commercial finite element code  (Comsol Multiphysics
\cite{comsol}) for the calculation of the effective indices. The
code uses
 an adaptive mesh algorithm for
generating the mesh. The meshes consist of 60,000-250,000
quadratical Lagrange elements. The core mode is simulated by
considering a cross section with 12-25 rings of air holes: 12 rings
for $d/\Lambda = 0.80-0.95$, 15 rings for $d/\Lambda = 0.50-0.75$,
and 25 rings for $d/\Lambda = 0.15-0.45$.  Outside the rings of air
holes there is a layer of air $0.5 \Lambda$ thick to reduce the
effects of a finite cladding.  The large number of rings of air
holes is chosen such that effects of a finite cladding is small.
Otherwise reflections may arise from both the silica-air interface
and the metallic boundary outside the air layer. This induces an
error on the effective index. In our simulation the number of rings
was sufficiently large  that the relative error on the effective
indices is smaller than $10^{-4}$. Large holes give a strong
confinement of the core mode, and obviously fewer rings are needed
compared to small holes.

We generally use only data where $n_\textrm{co} - n_\textrm{FSM}>
10^{-4}$ such that the error is negligible. This also makes physical
sense, since the PCF is only weakly guiding below this limit.

 For the fundamental space
filling mode we use the Bloch theorem and we need only to simulate a
single unit cell containing only one hole. The unit cell is spanned
by the lattice vectors seen in Fig.~(\ref{fig:analyte}) of an
infinite triangular cladding structure. We impose periodic boundary
conditions on the unit cell.

In our simulations we solve Eq.~(\ref{eq:eigenvalue}) for the
propagation constants, $\beta$, for the modes of the fiber at a
fixed frequency, $\omega$, which enables us to include material
dispersion in a consistent manner. For the refractive indices of the
materials we use a Sellmeier expression for the silica and use the
refractive index of water given in Ref.~\cite{water}.
All plots in the following are shown as function of the wavelength
relative to the pitch, but at a fixed wavelength. If the material
dispersion is neglected then space and time in Maxwell's equation
may be scaled with the pitch of the PCF. However, in our study
 the material dispersion cannot be neglected.
We thus choose a wavelength $\lambda = 1.0\mu$m.


The grating period and therefore the beat length for LPGs can be
determined from the effective indices using Eq.~(\ref{eq:res_LPG}).
The beat length varies significantly with the hole size and pitch,
 as shown in Fig. \ref{fig:lg}. The resonance condition,
Eq.~(\ref{eq:res_LPG}), can be satisfied over a range of wavelengths
if $\partial_\lambda \Lambda_\textrm{G} = 0$. For these values the
FWHM of an LPG is very large.

The between the curves of in Fig. \ref{fig:lg} the gray area
indicates $\bar n_\textrm{g} < 0$ and the white area  $\bar
n_\textrm{g} > 0$. The boundary between these two areas indicates
$\bar n_\textrm{g} = 0$.

When the slopes of the curves in Fig. \ref{fig:lg} are parallel with
the normalized wavelength axis the resonance condition
Eq.~\ref{eq:res_BG} is nearly matched for multiple normalized
wavelengths, $\lambda/\Lambda$, we may thus expect that the
wavelength FWHM is wide. By Eq.~(\ref{eq:fwhm}) the FWHM is widest
when $\bar n_\textrm{g} = 0$, indicated by the boundary of the gray
area. It may seem surprising that the  $\bar n_\textrm{g} = 0$ does
not coincide with zero slope of the curves. The curves in
Fig.~\ref{fig:lg}, however, are evaluated for a fixed wavelength
with a variable pitch, and not for a fixed pitch with variable
wavelength. The difference in normalized wavelength for group index
matching and for zero slope of the curves is caused by material
dispersion of silica.

For BGs $\bar n_\textrm{f}$ in the resonance condition
Eq.~(\ref{eq:res_BG}) only varies a few percent for all wavelengths,
pitch values, and hole sizes. Thus the dispersion effects on the
resonance condition are small. Correspondingly, it can shown that
dispersion effects on the wavelength FWHM is also small, because the
group index mismatch, $\bar n_{\textrm{g}}$, in Eq.~(\ref{eq:fwhm})
only varies slightly ($<$ 12\%) for index guiding PCFs, since
$n_\textrm{co}$ always dominates over its derivative in the group
index, $n_\textrm{g,co}= n_\textrm{co}-\lambda\partial_\lambda
n_\textrm{co}$.

There is a fundamental difference between PCFs and standard optical
fibers is that a PCF consists of two, often very different,
materials. Standard optical fibers also consist of two materials:
pure silica and a core of doped silica. In practice, however, the
material parameters for the doped silica may often, to a good
approximation, be taken to be equal to those of pure silica. For a
PCF, e.g. the strain optic response of silica is completely
different from that of air, which is obviously zero for our
applications.

To correctly handle a two material fiber one must determine the
field energy intensity fraction, $f_u$, inside the holes of the PCF.
The field energy fraction for the base material is then, $1-f_u$. It
is not possible to simply add a contribution, e.g. stress optic,
directly to the resonance condition, Eq.~(\ref{eq:res_BG}). Instead
every contribution must be weighted with the fraction of field
intensity in the corresponding material, as shown in the appendix.

It is possible to make some generalizations about field energy
fraction in the holes, $f_u$, for index guiding fibers. In the short
wavelength limit where the wavelength is much smaller than the pitch
of the structure, $\lambda\ll \Lambda$, the field is confined to the
high index base material. In the long wavelength limit, $\lambda\gg
\Lambda$, the changes in the field are small over a length of the
pitch. This imply
\begin{eqnarray}
&&
\begin{array}{rcl} f_u \to 0& \textrm{for} &\lambda/\Lambda  \to
0,
\end{array}\\
&&
\begin{array}{rcl} f_u \to f& \textrm{for} &\lambda/\Lambda \to
\infty,
\end{array}
\end{eqnarray}
where $f$ is the air filling fraction, i.e. the fraction of area
covered by the holes in the cross section of the fiber. The
approximation $f_u = 0$ is natural as a first approximation since
the field intensity in the air holes $f_u$ is generally much smaller
than the field intensity in the base material. But PCFs can
 possibly have air filling fractions of almost unity, implying that
also $f_u$ can be almost unity, and therefore the effect can not be
neglected.

\begin{figure}[b!]
\begin{center}
\includegraphics[angle = 0, width =0.45\textwidth]{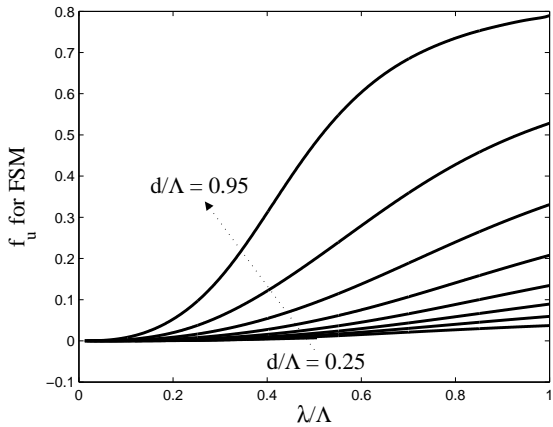}
\end{center}
\caption{Field energy intensity fraction in the air filled holes for
the fundamental space filling mode. The lines indicate different
values of the hole diameter relative to the pitch: 0.15 to 0.95 in
steps of 0.10}\label{fig:fufsm_air}
\end{figure}

\begin{figure}[b!]
\begin{center}
\includegraphics[angle = 0, width =0.45\textwidth]{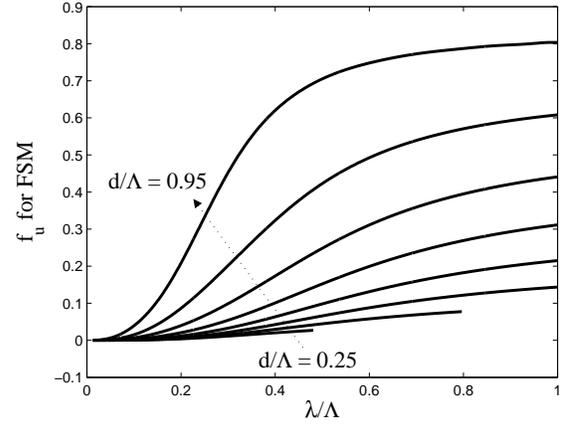}
\end{center}
\caption{Field energy intensity fraction in the water filled holes
for the fundamental space filling mode. The lines indicate different
values of the hole diameter relative to the pitch: 0.15 to 0.95 in
steps of 0.10 increasing from below}\label{fig:fufsm_h2o}
\end{figure}

\begin{figure}[b!]
\begin{center}
\includegraphics[angle = 0, width =0.45\textwidth]{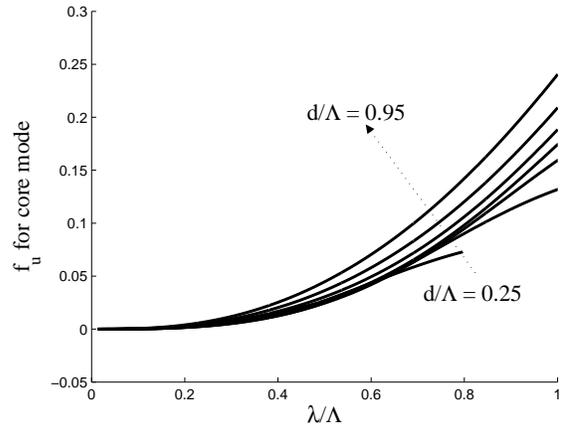}
\end{center}
\caption{Field energy intensity fraction in the water filled holes
for the core mode. The lines indicate different values of the hole
diameter relative to the pitch: 0.15 to 0.95 in steps of 0.10
increasing from below}\label{fig:fuco_h2o}
\end{figure}

 We have shown $f_u$ for the FSM in Fig.~\ref{fig:fufsm_air} for air filled holes ($n_h=1.0$). $f_u$ for the core
mode of a PCF with water filled holes ($n_h\simeq 1.33$) is shown in
Fig.~\ref{fig:fuco_h20}) and the corresponding FSM is shown in Fig.
\ref{fig:fufsm_air}. Clearly
 $f_{u}$ is larger for the FSM than for the core mode, and it increases
 with increasing wavelength and increasing hole diameters $d/\Lambda$.
The $f_{u}$ also increases when the index contrast between the hole
and base material is decreased. Hence $f_{u}$ is larger for water
filled PCFs than for air filled.


\subsection{Small and large core PCF}
In the following we refer to two PCFs to as the large core and the
small core PCF. The structure parameters for the small core PCF is
$(d/\Lambda, \Lambda) = (0.45, 7\mu m)$ 2) and for the large core
PCF $(d/\Lambda, \Lambda) = (0.75, 2\mu m)$.  We choose the length
of the gratings to be 30 mm for LPGs with additionally $\kappa =
4/L$ for BGs. The sensitivity, $\gamma$, is independent of these
parameters. The quality factor, $Q$, depends inversely proportional
on the length of the grating, when assuming constant $\kappa L$. For
BGs the quality factor also depends inversely proportional to the
coupling constant, $\kappa$.

\section{Refractive index sensing and biosensing}
\subsection{Refractive index sensing}
Refractive index sensors can be used in the application of
continuous monitoring of gasses in the oil industry.  The gas flows
through the PCF is detected by measuring the overall refractive
index of the hole contents. Optical fibers have the advantage that
there are no electric currents thus avoiding completely the
possibility of generating electric sparks.

The sensitivity for refractive index sensing is also used for
benchmarking the sensitivity of evanescent wave biosensors, which
will be studied in the next section.

Alternatively, a refractive index sensor may be used as tunable
grating if the holes are filled with a substance whose refractive
index can be controlled \cite{kerbage2003, rindorf2006_7,
sorensen2006}. Tunable gratings may have applications within
modulation of optical signals.

 Infiltrating substances  into PCF-LPGs  that are absorbing at the resonant wavelength
requires
 special considerations, as discussed by Daxhelet et al.
\cite{daxhelet2003}. In brief the attenuation induced by the
absorption of the substance becomes significant if the modes are
significantly attenuated over the length of a period of the grating.
The interference of modes in the grating becomes weakened and the
resonance dip is broadened. Since the grating period of an LPG is
much longer than that of a BG, the LPG can be considerably affected
by absorbtion, while the BG is largely unaffected even by a high
absorbtion.

The change in the  effective index of a mode induced by a change of
refractive index of the hole contents is proportional to the
fraction of the time averaged field energy intensity ($u(\rrr_\bot)
= \frac 1 2\varepsilon(\rrr_\bot)|\EEE(\rrr_\bot)|^2$) inside the
holes. For BGs the change in $\bar n_\textrm{f} = 2n_\textrm{co}$ is
(see appendix for derivation)
\begin{eqnarray}
\frac{\partial \bar n_\textrm{f}}{\partial n_h} = 2\,
\frac{n_\textrm{g,co}}{n_h} f_{u,\textrm{co}}
 ,
\label{eq:bgrefrac}
\end{eqnarray}
where $\Delta n_h$ is the change in the refractive index, $n_h$, of
the content in the hole. $f_{u,\textrm{co}}$ is the fraction of the
field energy intensity of the core mode inside the holes. The
sensitivity is  also proportional to the group index
$n_\textrm{g,co}$.
 For LPGs the change in $\bar n_\textrm{f} = n_\textrm{co}-n_\textrm{cl}$ is
\begin{eqnarray}
\frac{\partial \bar n_\textrm{f}}{\partial n_h} =
\frac{n_\textrm{g,co}}{n_h}f_{u,\textrm{co}}
 -
\frac{n_\textrm{g,cl}}{n_h} f_{u,\textrm{cl}}.
\label{eq:lpgrefrac}
\end{eqnarray} It is clear from
Eqs.~(\ref{eq:gamma},\ref{eq:q}) that the sensitivity and the
detectability are both largely determined by $f_u$. For BGs $f_u$
should be as large as possible.

For BGs the group index is close to the effective index,
$n_\textrm{g,co}  =  n_\textrm{co} - \lambda \partial_\lambda
n_\textrm{co} \approx n_\textrm{co}$ since the derivative is
negligible as seen in Fig.~\ref{fig:neffs}. Thus for BGs the
sensitivity is determined by $f_{u,\textrm{co}}$. In
Fig.~\ref{fig:fuco_h2o} we realize that a large  $f_{u,
\textrm{co}}$ implies a small pitch and a large hole diameter. For
BGs the size of $Q$ is largely determined by the pitch. $Q$ also
increases with decreasing hole size, but to a lesser extent.

\begin{figure}[b!]
\begin{center}
\includegraphics[angle = 0, width =0.45\textwidth]{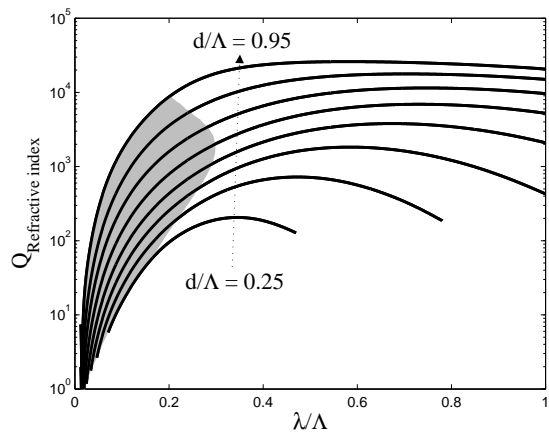}
\end{center}
\caption{Quality factor $Q_{\rm RI}$ for LPG refractive index
sensing. The lines indicate different values of the hole diameter
relative to the pitch: 0.25 to 0.95 in steps of 0.10. The gray area
indicates negative group index mismatch $\bar n _\textrm{g} < 0$.
The wavelength is 1 $\mu$m, the length of the gratings is $L$ = 30
mm, and $|\kappa L | = \pi/(2L)$}\label{fig:lpg_nr_Q}
\end{figure}

In order to have a high $Q$ for LPGs the difference
$n_\textrm{g,co}f_{u,\textrm{co}}
 -
n_\textrm{g,cl} f_{u, \textrm{cl}}$ should be large. Comparing
Figs.~\ref{fig:fuco_h2o} and \ref{fig:fufsm_h2o} we realize that
this also implies a small pitch and a large hole diameter. But in
this case the difference falls off beyond a certain point in the
wavelength pitch ratio. This is because $f_{u, \textrm{co}}$ and
$f_{u, \textrm{cl}}$ approaches the same limit for large
wavelengths, and thus the $Q$ decreases, as seen in Fig.
\ref{fig:lpg_nr_Q}.

 In Fig. \ref{fig:lpg_nr_Q}
 there is also indicated the group index matched line for LPG where
 $ \bar n_\textrm{f} = n_\textrm{g,co} -n_\textrm{g,cl}=0$.
Along this line the peak of the LPG is very wide by
Eq.~(\ref{eq:fwhm}). On the left side of this line the shift in
wavelength are positive for a positive $\Delta n_h$. On the right
hand side the shift in wavelength is negative for positive $\Delta
n_h$. The reason for this, is that $\bar n_\textrm{g}$ enters in
Eq.~(\ref{eq:gamma}) in the denominator, and thus its sign
determines the sign of the wavelength shift. For BGs group index
matching $\bar n_\textrm{f} =
2n_\textrm{co}-2\lambda\partial_\lambda n_\textrm{co} =0$ does not
occur since for the considered index guiding PCFs the effective
index is always larger than its derivative.

\begin{figure}[b!]
\begin{center}
\includegraphics[angle = 0, width =0.45\textwidth]{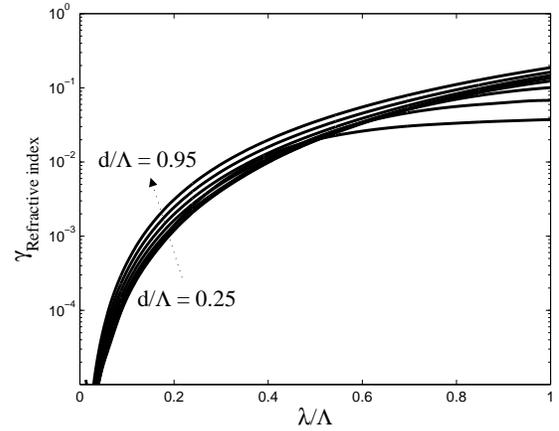}
\end{center}
\caption{$\gamma_{\rm RI}$ for BG refractive index sensing. The
lines indicate different values of the hole diameter relative to the
pitch: 0.15 to 0.95 in steps of 0.10. The wavelength is 1 $\mu$m,
the length of the gratings is $L$ = 30 mm, and $|\kappa L | =
4$}\label{fig:bg_nr_sens}
\end{figure}

For the sensitivity of BGs Eq.~(\ref{eq:bgrefrac}) is inserted in
Eq.~(\ref{eq:gamma}), and plotted for the wavelength 1.0 $\mu$m, in
Fig.~\ref{fig:bg_nr_sens}. For LPGs Eq.~(\ref{eq:lpgrefrac}) is
inserted in the expression for the $Q$, Eq.~(\ref{eq:q}) and plotted
for the wavelength 1.0 $\mu$m, Fig.~\ref{fig:lpg_nr_Q}.

To give some specific details we choose a typical large mode area,
endlessly single mode PCF with
 structure values: $\Lambda \simeq 7\mu$m and
$d/\Lambda\simeq 0.45$. This gives a core diameter of 10 $\mu$m. We
choose the length of the grating to be 30 mm, and the coupling
strength is $\kappa = \pi/(2L)$. We choose different wavelengths for
both LPGs and BGs: 600, 900, and 1550 nm. The data is presented in
Table \ref{table:refrac_lin}. The sensitivity increases with
increasing wavelength since the evanescent wave is increasing inside
the holes, corresponding to an increasing $f_u$. The sensitivity,
$\gamma$,
 for LPGs is four orders of magnitude larger than the sensitivity
 for BGs. But taking the FWHM of the resonance peak into account, the $Q$ is
 only slightly larger for LPGs than for BGs.

 The data for a small core PCF
are presented in Table \ref{table:refrac_nl}. The small core PCF has
significantly higher sensitivity than the large core PCF. Notice
that the sensitivity for the small core has opposite sign to the
large core PCF. Also $Q$ \emph{decreases} with increasing
wavelength. The shift is negative because the PCF's structure
parameters places the PCF on the left of the group index matching
line $\bar n_\textrm{g} = 0$ indicated in Fig. \ref{fig:lpg_nr_Q}.

The minimum detectable change in refractive index for LPGs can be
calculated. For L = 30 mm, we get a maximal $Q \sim 10^4$. Taking
that the minimal detectable wavelength is
$\lambda_\textrm{FWHM}/100$ we get a maximal sensitivity of $\Delta
n_\textrm{h,min} = \frac 1 {100} \frac 1 {Q} \sim 10^{-6}$. The same
calculation for a BG with $Q \sim 10^{2}$ gives $\Delta
n_\textrm{h,min}  \sim 10^{-4}$.

Both LPGs \cite{rindorf2006_7} and BGs \cite{phanhuy2006} have been
used for measuring the refractive index of a liquid analyte inside
the holes of the PCF. Comparing with experiment for LPGs
\cite{rindorf2006_7} the PCF has parameters $d/\Lambda = 0.47,
\Lambda = 7.2 \mu$m, and the LPG has $\lambda_\textrm{r} \simeq
845$nm and $L$ = 18.2 mm. The minimal detectable wavelength shift is
$\Delta \lambda _\textrm{min} \approx 1$ nm.  Theoretically the
parameters give $Q_{\rm RI}$ = 52 (no units) and
$\lambda_\textrm{FWHM} \simeq 45$ nm which agrees with the
experimental value $\lambda_\textrm{FWHM} \approx 45$ nm. The
minimal sensitivity given the mentioned minimal detectable
wavelength  is theoretically $\Delta n_\textrm{h,min} \approx
4.3\times10^{-3}$ which is in agreement with the experimental value
$\Delta n_\textrm{h,\textrm{min}} \approx 10^{-4}$. The shift in
resonances wavelength as function of sensitivity is experimentally
$\Delta\lambda_\textrm{r}/\Delta n_\textrm{h}$ = 7500 nm but gives
theoretically a lower value $\Delta\lambda_\textrm{r}/\Delta
n_\textrm{h}$ = 2300 nm.

\begin{widetext}
\begin{center}
\begin{table}
\begin{tabular}{|c|c|c|c|c|c|c|c|}
\hline {\bf large-core PCF} & \multicolumn{3}{|c|}{\bf Refractive index sensing  }&\multicolumn{3}{|c|}{\bf Biosensing }&\\
\cline{2-7}
 & $\gamma_\textrm{RI}$ & $Q_\textrm{RI}$ & $\Delta\lambda/\Delta n_\textrm{h}$ [nm] & $\gamma_\textrm{Bio}$ [1/nm] & $Q_\textrm{Bio}$ [1/nm] & $\Delta\lambda/t_\textrm{Bio}$ & $\lambda_\textrm{\scriptsize FWHM}$ \\
\hline LPG, 600 nm & 1.6 & 43 &  980 nm & 2.4$\times 10^{\rm \scriptsize 3}$ & 63$\times 10^{\rm \scriptsize -3}$ & 1.4 nm/nm & 23 nm \\
\hline LPG, 900 nm & 3.2 & 88 & 2700 nm & 3.0 $\times 10^{\rm \scriptsize 3}$ & 85$\times 10^{\rm \scriptsize -3}$ & 2.7 nm/nm & 32 nm \\
\hline LPG,1550 nm & 14 & 210 & 22$\times 10^{\rm \scriptsize 3}$ nm & 8.1 $\times 10^{\rm \scriptsize -3}$ & 120$\times 10^{\rm \scriptsize -3}$ & 12 nm/nm & 109 nm \\
\hline BG, 600 nm & 0.10$\times 10^{\rm \scriptsize -3}$ & 5.8 & 0.060 nm &3.0$\times 10^{\rm \scriptsize -7}$  & 17$\times 10^{\rm \scriptsize -3}$ & 0.18 pm/nm & 10 pm \\
\hline BG, 900 nm & 0.34$\times 10^{\rm \scriptsize -3}$ & 13 & 0.30 nm & 6.5$\times 10^{\rm \scriptsize -7}$ & 25$\times 10^{\rm \scriptsize -3} $ & 0.59 pm/nm & 24 pm \\
\hline BG, 1550 nm &1.6$\times 10^{\rm \scriptsize -3}$ & 35 &  2.5 nm & 19$\times 10^{\rm \scriptsize -7}$ & 41$\times 10^{\rm \scriptsize -3} $ & 2.8 pm/nm &  70 pm\\
\hline\hline {\bf small-core PCF} & \multicolumn{3}{|c|}{\bf Refractive index sensing}&\multicolumn{3}{|c|}{\bf Biosensing } &\\
\cline{2-7}
 & $\gamma_\textrm{RI}$ & $Q_\textrm{RI}$ & $\Delta\lambda/\Delta n_\textrm{h}$ & $\gamma_\textrm{Bio}$ [1/nm] & $Q_\textrm{Bio}$ [1/nm] & $\Delta\lambda/t_\textrm{Bio}$ & $\lambda_\textrm{\scriptsize FWHM}$ \\
\hline LPG, 600 nm &  73  & 6500 & 44$\times 10^{\rm \scriptsize 3}$ nm & 89$\times 10^{\rm \scriptsize -3}$ & 7.9 & -54 nm/nm & 6.8 nm \\
\hline LPG, 900 nm & 11 & 8600 &  9.8$\times 10^{\rm \scriptsize 3}$ nm & 7.7 $\times 10^{\rm \scriptsize -3}$ & 6.2 & -7.0 nm/nm & 1.1 nm \\
\hline LPG, 1550 nm & 6.4 & 7400 &  9.9$\times 10^{\rm \scriptsize 3}$ nm & 2.7 $\times 10^{\rm \scriptsize -3}$ & 3.0 & -4.1 nm/nm & 1.3 nm \\
\hline BG, 600 nm & 6.0$\times 10^{-3}$ & 350 & 3.6 nm & 17$\times 10^{\rm \scriptsize -6}$ & 1.0 & 10 pm/nm  & 10 pm \\
\hline BG, 900 nm & 18$\times 10^{-3}$ & 690 & 16 nm &  35$\times 10^{\rm \scriptsize -6}$ & 1.33 & 31 pm/nm  & 23 pm \\
\hline BG, 1550 nm & 74$\times 10^{-3}$  & 1700 & 115 nm & 83$\times 10^{\rm \scriptsize -6}$ & 1.9 & 130 pm/nm  & 70 pm\\
\hline
\end{tabular}
\caption{Comparison of sensitivity and the quality factor for PCF
biosensors and refractive index sensors for both LPGs and BGs  at
different wavelengths for two different PCFs: a large core of 10
$\mu$m and a small core PCF of 1.5 $\mu$m.
}\label{table:bio_lin}\label{table:bio_nl}\label{table:refrac_lin}\label{table:refrac_nl}
\end{table}
\end{center}
\end{widetext}

 For BGs we compare with Huy
et al. \cite{phanhuy2006}. They have a 6 hole fiber with  $d
\approx$ 15 $\mu$m and pitch $\approx$ 15.8 $\mu$m. This gives
theoretically $\lambda_\textrm{FWHM} \simeq 117$ pm in agreement
with the experimental $\lambda_\textrm{FWHM} \approx 150$ pm.
Theoretically we have $\Delta\lambda_\textrm{r}/\Delta n_\textrm{h}
\approx 0.59$ nm and taking their a minimal detectable wavelength
shift of $\Delta \lambda _\textrm{min} = 1.0$ pm, we find $\Delta
n_\textrm{h,\textrm{min}} \approx 1.7\times 10^{-3}$ which is in
agreement with the experimental value $\Delta
n_\textrm{h,\textrm{min}} \approx 4 \times 10^{-3}$. Their numerical
simulations gives a minimal detectable shift which is a magnitude
off the experimental value.

\subsection{Label-free biosensor}
Biosensors in PCFs have seen some interesting developments in recent
times. The PCF biosensors can be made to detect both DNA, antibodies
or antigens, and the PCF can be made of both silica and polymer
materials. The experiments have been conducted using labeled
molecules, i.e. target biomolecules have a fluorescent tag, that can
be detected using fluorescence or absorbance microscopy. To be
really interesting the measurements must be label-free. This can be
done with PCF-LPG using the evanescent-wave sensing principle. This
principle is used in the widespread surface-plasmon resonance
biomolecule interaction analysis instruments.

The evanescent-wave principle measures the increase of refractive
index at a surface due to the presence of biomolecules. The
experiments are typically carried out in an aqueous environment.
 Water has refractive index
$\sim 1.33 $ and biomolecules have $\sim 1.45-1.48$. A given
biomolecule may be tested for its affinity to another biomolecule,
by immobilizing the first biomolecule onto the surface
(functionalization), and then infiltrating the second molecule into
the PCF. By a wash any biomolecules that have not been immobilized
may be flushed out. The immobilized biomolecules will constitute an
increase in the layer of molecules on the surface. The layer is 1-10
nm thick depending on the application.

The film of molecules on the surface may not constitute a
homogeneous layer. But since the wavelength is much longer than the
thickness of the layer (1 micron $\gg$ 10 nm), the sensor measures
the thickness of the biofilm \emph{on average}. The sensor is thus
largely insensitive to inhomogeneous distributions of molecules on
the surface.
 For benchmarking we assume that the layer of molecules has a
homogeneous thickness $t_\textrm{Bio}$ and has a refractive index of
$1.45$. This refractive index is to a good approximation identical
to silica: $n_\textrm{r,\textrm{Bio}} \simeq n_{\textrm{SiO}_2}$.
Thus a layer of biomolecues $t_\textrm{Bio}$ thick corresponds to
decrease of $\Delta d = - t_\textrm{Bio}/2$ in the hole diameter. To
calculate the change in effective index due to a layer of molecules
we thus find
\begin{eqnarray}
\frac{\partial  n_\textrm{eff}}{\partial t_\textrm{Bio}} = -2\frac{
\partial n_\textrm{eff}}{\partial d},
\end{eqnarray}
where $d$ is the hole diameter. The change in $\bar n_\textrm{f}$
for LPGs and BGs is
\begin{eqnarray}
\frac{ \partial  \bar n_\textrm{f}}{\partial t_\textrm{Bio}} &=&
\left\{
\begin{array}{cl}
- 2 \big( \frac{\partial  n_\textrm{co}}{\partial d}-\frac{\partial
n_\textrm{cl}}{\partial d}\big) ,
 & \qquad \textrm{LPG},\\
-  4\frac{\partial  n_\textrm{co}}{\partial d},
 & \qquad \textrm{BG}. \label{eq:deltan_bio}
\end{array}\right.
\end{eqnarray}

\begin{figure}[b!]
\begin{center}
\includegraphics[angle = 0, width =0.45\textwidth]{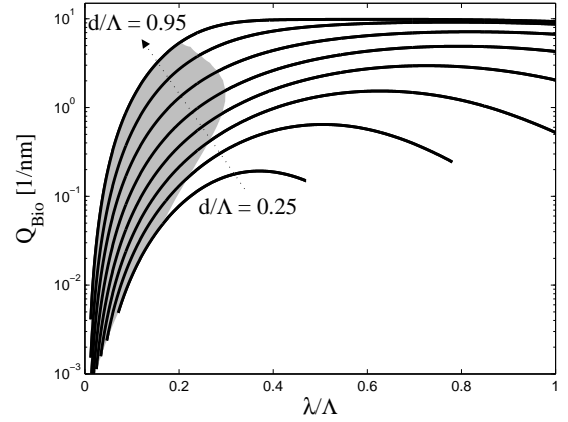}
\end{center}
\caption{Quality factor, $Q_{\rm Bio}$, for LPG biosensing. The
lines indicate different values of the hole diameter relative to the
pitch: 0.25 to 0.95 in steps of 0.10. The gray area indicates
negative group index mismatch $\bar n _\textrm{g} < 0$. The
wavelength is 1 $\mu$m, the length of the gratings is $L$ = 30 mm,
and $|\kappa L | = \pi/(2L)$}\label{fig:lpg_bio_Q}
\end{figure}

\begin{figure}[b!]
\begin{center}
\includegraphics[angle = 0, width =0.45\textwidth]{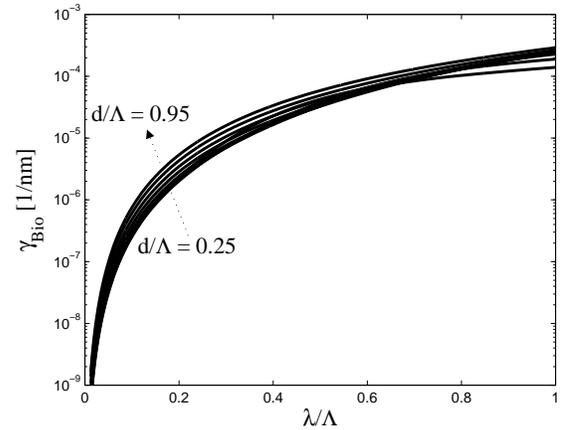}
\end{center}
\caption{Sensitivity, $\gamma_{\rm Bio}$, for BG biosensing. The
lines indicate different values of the hole diameter relative to the
pitch: 0.15 to 0.95 in steps of 0.10. The wavelength is 1 $\mu$m,
the length of the gratings is $L$ = 30 mm, and $|\kappa L | =
4$}\label{fig:bg_bio_sens}
\end{figure}

In Figs. \ref{fig:lpg_bio_Q} and \ref{fig:bg_bio_sens} it is seen
that $Q_{\rm Bio}$ and $\gamma_{\rm Bio}$ are largest for small
pitches and large wavelengths. For BGs the $\gamma_{\rm Bio}$ is
largely governed by the pitch. For LPGs a large benefit comes from
using large holes instead of small holes, whereas the pitch is less
important.

For BGs Eq.~(\ref{eq:deltan_bio}) is inserted in the expression for
the sensitivity, Eq.~(\ref{eq:gamma}) and plotted for the wavelength
1.0 $\mu$m, Fig.~\ref{fig:bg_bio_sens}. For LPGs
Eq.~(\ref{eq:deltan_bio}) is inserted in the expression for the $Q$,
Eq.~(\ref{eq:q}) and plotted for the wavelength 1.0 $\mu$m,
Fig.~\ref{fig:lpg_bio_Q}.

To compare with experiments we take the PCF structural values from
\cite{rindorf2006_7}: $d/\Lambda = 0.47$, $\Lambda = 7.2\mu m$. The
wavelength was $\approx 845$ nm. This gives  $\Delta \lambda = 2.3
\times t_{\rm Bio}$, $Q = 53\times 10^{\rm \scriptsize -3}$ (no
units), and $\gamma = 2.8\times 10^{\rm \scriptsize -3}/$nm. Thus a
1 nm thick layer of biomolecules gives a 2.3 nm shift in resonant
wavelength. It should be noted that the paper \cite{rindorf2006_7}
gives a value (1.4 nm/nm) that is $\sim 50$\% more. This is because
the effective indices were calculated using a different numerical
method, and material dispersion was not included in the simulations.

In comparison the 6 holes PCF of Huy et al. \cite{phanhuy2006} (d
$\approx$ 15 $\mu$m, pitch $\approx$ 15.8$\mu$m)  gives $\Delta
\lambda = 0.73\times 10^{\rm \scriptsize -3} \times t_{\rm Bio}$, $Q
= 6.2\times 10^{\rm \scriptsize -3}/$nm, and $\gamma = 0.47\times
10^{\rm \scriptsize -6}/$nm at $\lambda_\textrm{r} \simeq 1555$nm.
Thus a 1 nm thick layer of biomolecules gives a 0.73 pm shift in
resonant wavelength.

In general, both the sensitivity and the quality factor for BGs is
monotonically increasing with increasing wavelength. They also
increase with increasing hole size as seen in
Table~\ref{table:bio_lin}. The general behavior for the both the
sensitivity and the quality factor for LPGs is more complicated. The
sensitivity is strongly influenced by the group index matching, and
the resonant wavelength shifts can change sign depending on the PCF
structure parameters. The quality factor, $Q$, is high for large air
holes, and big normalized wavelength $\lambda/\Lambda$.

In conclusion, the sensitivity of label-free PCF grating biosensors
follows roughly the tendencies of the refractive index sensors. The
sensitivity varies several orders of magnitude in sensitivity and
quality factor with the PCF structure parameter. BGs have a high
sensitivity when the pitch is small compared with the wavelength.
Thus the optimal BG biosensor sensor is a small core fiber. For LPGs
the quality factor is high when the air holes are large, and are
less influenced by wavelength to pitch ratio, except in the short
limit where the quality goes to zero. The optimal LPG biosensor or
refractive index sensor thus has very large holes and a pitch about
three times or less the wavelength.

\section{Temperature and strain sensing}
\subsection{Temperature }
Sensitivity to temperature is often an unwanted source of error in
fiber gratings. However, fiber gratings may also be employed as high
temperature sensors since silica is stable to high temperatures. If
the susceptibility to temperature is an unwanted effect it may also
be remedied by using two multiplexed fiber gratings each with a
different linear temperature response. In the following we will
study the temperature response of air filled silica PCFs although
the theory is also applicable to other materials.

Temperature  changes both the material refractive indices of the PCF
through the thermo-optic effect and also by expansion of the PCF
structure. The thermo-optic coefficient for the  refractive indexof
silica glass is $C^n_T \simeq 9 \times 10^{-6}$/K and the thermal
expansion coefficient is $\alpha \simeq 0.55\times 10^{-6}$/K. Since
the perturbation of the PCF is sufficiently small we may expand the
effective index to first order in the temperature. Since both the
pitch, length and the refractive index of the silica are dependent
on silica, we use the partial derivatives of the effective
refractive index (the chain rule)
\begin{eqnarray} \frac{\partial n_{\rm eff}}{\partial T}\Big|_{\lambda_\textrm{r}}
&=& \frac{\ud n_{\rm eff}\big(L(T), \Lambda(T), n_r(T)\big)}{\ud T}
\nonumber
\\&=& \frac{\partial n_{\rm eff}}{\partial L}\frac{\partial L}{\partial T}
  + \frac{\partial n_{\rm eff}}{\partial
\Lambda}\frac{\partial \Lambda}{\partial T} + \frac{\partial n_{\rm
eff}}{\partial n_{\rm b}}\frac{\partial n_{\rm b}}{\partial T} .
\nonumber\\\label{eq:deltaneff_T}
\end{eqnarray}
We readily identify the thermal expansion$\frac{\partial
\Lambda}{\partial T}
 =  \alpha\Lambda $ and $\frac{\partial L}{\partial T}
 =\alpha L $.
The partial derivative of the effective index with respect to the
length can be evaluated according to following argumentation.  An
elongated PCF at a fixed effective index is equivalent to a PCF at
fixed length with an increased effective index.  Formulated in
mathematical terms this yields
\begin{eqnarray} \frac{\partial n_{\rm eff}}{\partial L}
= \frac{n_{\rm eff}}{L}.
\end{eqnarray}
In Eq.~(\ref{eq:deltaneff_T}) the partial derivative of the
effective index  with respect to the pitch (2nd term rhs) must be
evaluated numerically. The contribution from the thermo-optic effect
can be evaluated using
Eq.~(\ref{eq:deltaneff_refrac_app_base}). Eq.~(\ref{eq:deltaneff_T})
can then be written  for a BG as
\begin{eqnarray}
\frac{\partial \bar n_{\rm f}}{\partial T}\Big|_{\lambda_\textrm{r}}
&=& 2 \alpha \big(n_{\rm co} +\Lambda \frac{\partial n_{\rm
co}}{\partial \Lambda}\big) + \frac{C^n_{T}}{n_{\rm
b}}(1-f_\textrm{u,co})n_\textrm{g,co} \label{eq:bgtemp}.
\end{eqnarray}
For an LPG we get
\begin{eqnarray}\frac{\partial \bar n_{\rm f}}{\partial T}\Big|_{\lambda_\textrm{r}}
 &=&  \alpha \big( \bar
n_\textrm{f} + \Lambda \frac{\partial \bar n_\textrm{f}}{\partial
\Lambda}\big) \label{eq:lpgtemp}\\
&&\qquad+ \frac{C^n_{{\rm b},T}}{n_{\rm b}}\big[ (1-f_{u,
\textrm{co}})n_\textrm{g,co}-(1-f_{u,
\textrm{cl}})n_\textrm{g,cl}\big]\nonumber.
\end{eqnarray}
For generality we have kept a general refractive index of the base
material.

\begin{figure}[b!]
\begin{center}
\includegraphics[angle = 0, width =0.45\textwidth]{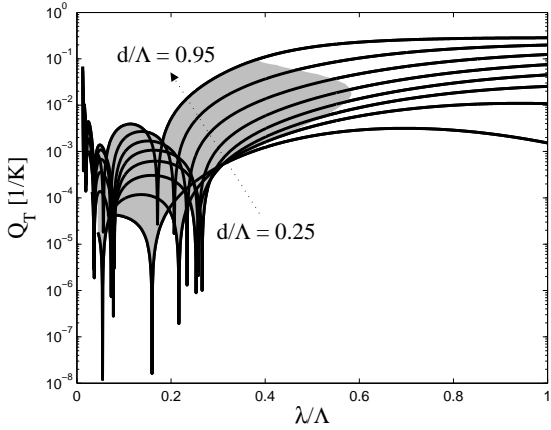}
\end{center}
\caption{Quality factor, $Q_T$, for LPG temperature sensing. The
lines indicate different values of the hole diameter relative to the
pitch: 0.25 to 0.95 in steps of 0.10. The gray area indicates
negative group index mismatch $\bar n _\textrm{g} < 0$. The
wavelength is 1 $\mu$m, the length of the gratings is $L$ = 30 mm,
and $|\kappa L | = \pi/(2L)$ }\label{fig:lpg_temp_Q}
\end{figure}

\begin{figure}[b!]
\begin{center}
\includegraphics[angle = 0, width =0.45\textwidth]{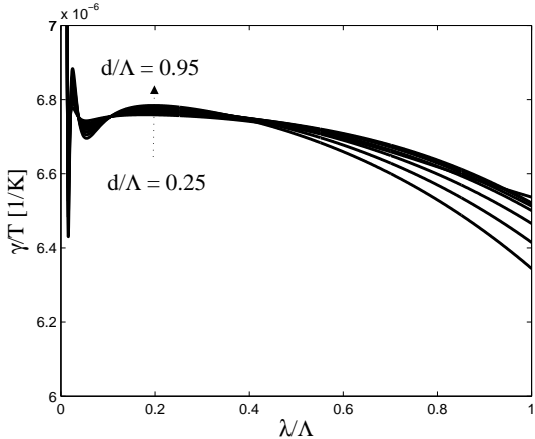}
\end{center}
\caption{Sensitivity, $\gamma_T$, for BG temperature sensing. The
lines indicate different values of the hole diameter relative to the
pitch: 0.15 to 0.95 in steps of 0.10. The wavelength is 1 $\mu$m,
the length of the gratings is $L$ = 30 mm, and $|\kappa L | =
4$}\label{fig:bg_temp_sens}
\end{figure}

 For BGs Eq.~(\ref{eq:bgtemp}) is inserted in the
expression for the sensitivity, Eq.~(\ref{eq:gamma}), and plotted
for the wavelength 1.0 $\mu$m, Fig.~\ref{fig:bg_temp_sens}. For LPGs
Eq.~(\ref{eq:lpgtemp}) is inserted in the expression for the $Q$,
Eq.~(\ref{eq:q}), and plotted for the wavelength 1.0 $\mu$m in
Fig.~\ref{fig:lpg_temp_Q}.


The sensitivity of BG PCF temperature sensors is almost constant for
 the structure parameters, $\Lambda$ and $d/\Lambda$,
Fig. \ref{fig:bg_temp_sens}. The $Q$ value for PCF-LPG temperature
sensors in contrast varies several orders of magnitude with the
structure parameters, $\Lambda$ and $d/\Lambda$, Fig.~
\ref{fig:lpg_temp_Q}. The dips seen on the left in the figures is
a change of sign of the wavelength shift.

In table \ref{table:temp_lin} we see that the resonant wavelength
shifts for both BGs and LPGs are of the order of 5pm/K, regardless
of PCF parameters. The quality factor is several orders of magnitude
smaller for LPGs compared with BGs. This can be attributed to the a
partial cancelation of the thermo-optic terms of the core and
cladding mode in Eq.~(\ref{eq:lpgtemp}).

We note that the sensitivities for the BGs are surprisingly constant
in Table \ref{table:temp_lin}. In the rhs of Eqs.~(\ref{eq:lpgtemp},
\ref{eq:bgtemp}) the terms \protect{$n_{\rm eff} + \Lambda
\frac{\partial n_{\rm eff}}{\partial \Lambda}$} and
$n_\textrm{g}(1-f_u)/{n_b}$ are roughly of the same order of
magnitude. The prefactor for the thermo-optic term is larger than
the prefactoc of thermal expansion term, $C_n^T /\alpha \approx 20$.
The temperature response of the grating is largely determined by the
thermo-optic. For BGs we may thus approximate $\gamma_T = 1/\lambda
\frac{\partial\lambda_{\rm r}}{\partial T} = C_n^T/n_\textrm{b}
\Delta T$. For silica PCFs $\gamma_T \simeq C_n^T/n_\textrm{b}
\simeq 0.62 \times 10.6/$K in good agreement with the values given
in tables \ref{table:temp_nl} and \ref{table:temp_lin}. We conclude
that sensitivity of BG temperature sensors is constant, within a few
\% margin, for all PCF structures and is dominated by the
thermo-optic coefficient relative to the refractive index of the
base material.

 To validate our theory we compare with experiments. Dobb et al.
\cite{dobb2006} characterized LPGs in PCFs inscribed with an
electric arc. For a PCF with structure parameters $\Lambda \simeq
7.1$ $ \mu$m, $d \simeq 4.0\pm0.4$ $\mu$m they found a shift of
resonance wavelength with temperature as 3.4 pm/K at the wavelength
1511 nm. Using the structure parameters in the theory yields 4.3
pm/K. For a PCF with structure parameters $\Lambda \simeq 8$ $
\mu$m, $d \simeq 3.7\mu$m, they found 2.2 pm/K at 1403 nm wavelength
and the theory gives 6.1 pm/K. Thus we have an agreement of the
right magnitude and right sign of the wavelength shift. It is seen
in Fig. \ref{fig:bg_temp_sens} that for their structure values,
$d/\Lambda \simeq 0.46, \lambda/\Lambda \simeq 0.18$  the quality
factor, $Q$, experiences a series of dips.  The dips correspond to
the terms in Eq.~(\ref{eq:lpgtemp}) cancel in the net contribution.
Both numerically, and experimentally, an estimation of the
sensitivity will be prone to small errors on contribution from each
term.

In an earlier experiment Humbert et al. \cite{humbert2003} used the
electric arc method to inscribe LPGs in a PCF with structure
parameters $\Lambda \simeq 4.0 \mu$m, $d \simeq 2.0\mu$m. They found
a value of 9 pm/K at 1590 nm, roughly half the theoretical value -23
pm/K, but with an opposite sign. To inscribe BGs Martelli et al.
used a PCF with a erbium-doped UV sensitive core
\cite{martelli2005}. Their structure parameters are roughly $\Lambda
\simeq 10 \mu$m, $d \simeq 2\mu$m.
 At a wavelength of 1535 nm the wavelength shift is
20 pm/K. The theory gives 10 pm/K for a pure silica PCF. An erbium
doped core is known to decrease the temperature sensitivity in
standard optical fibers.

BGs have also been inscribed in a small core fiber with the
structure parameters $d/\Lambda$ = 0.5 and  $\Lambda = 1.6 \mu$m
\cite{frazao2005}. The core was doped with germanium to make the
core UV-sensitive. They found an experimental sensitivity of 4.11
pm/K at a wavelength of 1508 nm. Our theory gives 9.9 pm/K. A
Germanium doped core is known to decrease the temperature
sensitivity in standard optical fibers.

To  enhance the temperature sensitivity in PCF the holes can be
filled with substances with a high thermo-optic coefficient. This
has been shown to increase temperature sensitivity
\cite{sorensen2006}. Eq.~(\ref{eq:lpgtemp}) can in this case be
modified to be valid by adding an additional temperature optic term
for the holes.


\begin{widetext}
\begin{center}
\begin{table}
\begin{tabular}{|c|c|c|c|c|c|c|c|}
\hline {\bf large-core PCF} & \multicolumn{3}{|c|}{\bf Temperature}&\multicolumn{3}{|c|}{\bf Strain}&\\
\cline{2-7}
 & $\gamma_\textrm{T}$ & $Q_\textrm{T}$  [1/K] & $\Delta\lambda/\Delta T$ [1/K]& $\gamma_{\varepsilon_s} $ & $Q_{\varepsilon_s}$  & $\Delta\lambda/{\varepsilon_s}$ & $\lambda_\textrm{\scriptsize FWHM}$ \\
\hline LPG, 600 nm & 3.9$\times 10^{\rm \scriptsize -6}$ & 14$\times 10^{\rm \scriptsize -5}$ & 2.3 pm &-1.2$\times 10^{\rm \scriptsize -6}$ & 4.5$\times 10^{\rm \scriptsize -5}$ & -0.74 pm & 17 nm \\
\hline LPG, 900 nm & 6.2$\times 10^{\rm \scriptsize -6}$ & 30$\times 10^{\rm \scriptsize -5}$ & 5.6 pm & -2.3$\times 10^{\rm \scriptsize -6}$ & 11$\times 10^{\rm \scriptsize -5}$ & -2.1 pm &19 nm \\
\hline LPG, 1550 nm & 1.7$\times 10^{\rm \scriptsize -6}$ & 10$\times 10^{\rm \scriptsize -5}$ & 27 pm & -2.1$\times 10^{\rm \scriptsize -6}$ & 13$\times 10^{\rm \scriptsize -5}$ & -3.3 pm & 26 nm \\
\hline BG, 600 nm & 6.7$\times 10^{\rm \scriptsize -6}$ & 0.39 & 4.0 pm & 0.82$\times 10^{\rm \scriptsize -6}$ & 0.047 & 0.49 pm &  10 pm \\
\hline BG, 900 nm & 6.8$\times 10^{\rm \scriptsize -6}$ & 0.26 & 6.1 pm & 0.82$\times 10^{\rm \scriptsize -6}$ & 0.031 & 0.73 pm &24 pm\\
\hline BG, 1550 nm & 6.8$\times 10^{\rm \scriptsize -6}$ & 0.15 & 11 pm & 0.81$\times 10^{\rm \scriptsize -6}$ & 0.018 & 1.3 pm &70 pm \\
\hline\hline {\bf small-core PCF} & \multicolumn{3}{|c|}{\bf Temperature}&\multicolumn{3}{|c|}{\bf Strain}&\\
\cline{2-7}
 & $\gamma_\textrm{T}$ & $Q_\textrm{T}$  [1/K] & $\Delta\lambda/\Delta T$ [1/K]& $\gamma_{\varepsilon_s} $ & $Q_{\varepsilon_s}$  & $\Delta\lambda/{\varepsilon_s}$ & $\lambda_\textrm{\scriptsize FWHM}$ \\
\hline LPG, 600 nm & 5.0$\times 10^{\rm \scriptsize -5}$ & 5.8$\times 10^{\rm \scriptsize -3}$ & -3.0 pm &-2.4$\times 10^{\rm \scriptsize -6}$ & 2.8$\times 10^{\rm \scriptsize -3}$ & -1.4 pm & 0.52 nm \\
\hline LPG, 900 nm & 4.0$\times 10^{\rm \scriptsize -5}$ & 23$\times 10^{\rm \scriptsize -3}$ & -36 pm & -4.2$\times 10^{\rm \scriptsize -6}$ & 2.4$\times 10^{\rm \scriptsize -3}$ & -2.0 pm &1.6 nm \\
\hline LPG, 1550 nm & 3.7$\times 10^{\rm \scriptsize -5}$ & 55$\times 10^{\rm \scriptsize -3}$ & 57 pm & 0.7$\times 10^{\rm \scriptsize -6}$ & 1.0$\times 10^{\rm \scriptsize -3}$ & 1.1 pm & 1.0 nm\\
\hline BG, 600 nm  & 6.7$\times 10^{\rm \scriptsize -6}$ & 0.39 & 4.0 pm & 0.81$\times 10^{\rm \scriptsize -6}$ & 0.047 & 0.48 pm & 10 pm \\
\hline BG, 900 nm  & 6.7$\times 10^{\rm \scriptsize -6}$ & 0.26 & 6.1 pm & 0.79$\times 10^{\rm \scriptsize -6}$ & 0.030 & 0.71 pm &  23 pm \\
\hline BG, 1550 nm & 6.6$\times 10^{\rm \scriptsize -6}$ & 0.15 & 10 pm & 0.75$\times 10^{\rm \scriptsize -6}$ & 0.017 & 1.2 pm &68 pm\\
\hline
\end{tabular}
\caption{Comparison of sensitivity and the quality factor for PCF
temperature and strain sensors for
both LPGs and BGs at different wavelengths for two different PCFs: a
large core of 10 $\mu$m and a small core PCF of 1.5 $\mu$m
}\label{table:temp_lin}\label{table:temp_nl}\label{table:strain_lin}\label{table:strain_nl}
\end{table}
\end{center}
\end{widetext}

\subsection{Strain}
Fiber optic strain sensors have found application within civil
engineering, where they provide a stable and sensitive measurement
of the strain in constructions. Strain is defined as the relative
elongation: $\varepsilon_s = \Delta L /L$ of a piece PCF of length
$L$. Often millistrain, $\mu\varepsilon_s = 10^6\varepsilon$, is
used instead of strain. When the PCF is elongated, the PCF contracts
transversely to the applied strain. This is expressed by Poisson's
ratio which formally is the ratio between lateral and tensile strain
with a negative sign. For silica glass Poisson's ratio is given by
$\nu = 0.17$. Strain also changes the refractive index of materials.
The strain-optic coefficient for silica is $C^n_\epsilon =
\eta_\epsilon = -0.26$ \cite{park2002}. Since strain changes the
length, pitch, and the material refractive indices we apply partial
derivatives (the chain rule) to the linearization of the effective
index
\begin{eqnarray}
\frac{\partial n_{\rm eff}}{\partial
\varepsilon_s}\Big|_{\lambda_\textrm{r}} &=& \frac{\ud
n_\textrm{eff}\big(L(\varepsilon_s),
\Lambda(\varepsilon_s),n_r(\varepsilon_s)\big)}{\ud \varepsilon_s}
\nonumber
\\&=& \frac{\partial n_\textrm{eff}}{\partial L}\frac{\partial L}{\partial \varepsilon_s}
 + \frac{\partial n_\textrm{eff}}{\partial
\Lambda}\frac{\partial \Lambda}{\partial \varepsilon_s} +
\frac{\partial n_\textrm{eff}}{\partial n_r}\frac{\partial
n_r}{\partial
\varepsilon_s}\nonumber. \\
\end{eqnarray}
Since $L(\varepsilon_s) = (1+\varepsilon_s)L$ we have $
\frac{\partial L}{\partial \varepsilon_s} = L$ and
$\Lambda(\varepsilon_s) = (1+\varepsilon_s)\Lambda$ gives
$\frac{\partial \Lambda}{\partial \varepsilon_s} = -\nu \Lambda$. An
increase in length of the PCF with a fixed effective index is
mathematically equivalent to a fixed length with an increase in
effective index,
\begin{eqnarray}
\frac{\partial n_\textrm{eff}}{\partial L} &=& \frac{
n_\textrm{eff}}{ L}.
\end{eqnarray}
 The derivative
of the effective index with respect to the refractive index of the
base material may be evaluated by inserting the strain-optic
coefficient in the perturbative formula,
Eq.~(\ref{eq:deltaneff_refrac_app_base}), given in the appendix. The
formula for BGs is
\begin{eqnarray}
\frac{\partial n_{\rm f}}{\partial
\varepsilon_s}\Big|_{\lambda_\textrm{r}} &=&
 2n_\textrm{co}  - \nu\Lambda\frac{\partial  2 n_\textrm{co}}{\partial
\Lambda} \nonumber \\
&&+ \frac{\eta_\varepsilon}{n_b} 2n_\textrm{g,co}(1-f_{u,
\textrm{co}}) \label{eq:bgstrain},
\end{eqnarray}
while for LPGs we obtain
\begin{eqnarray}
\frac{\partial \bar n_{\rm f}}{\partial
\varepsilon_s}\Big|_{\lambda_\textrm{r}}
 &=&
 (n_\textrm{co}-n_\textrm{cl})  - \nu\Lambda\frac{\partial  (n_\textrm{co}-n_\textrm{cl})}{\partial
\Lambda} \label{eq:lpgstrain}
 \\&&\hspace{-1.5cm}+ \frac{\eta_\varepsilon}{n_b}
\big( n_\textrm{g,co}(1-f_{u, \textrm{co}})-n_\textrm{g,cl}(1-f_{u,
\textrm{cl}})\big) \nonumber.
\end{eqnarray}

\begin{figure}[b!]
\begin{center}
\includegraphics[angle = 0, width =0.45\textwidth]{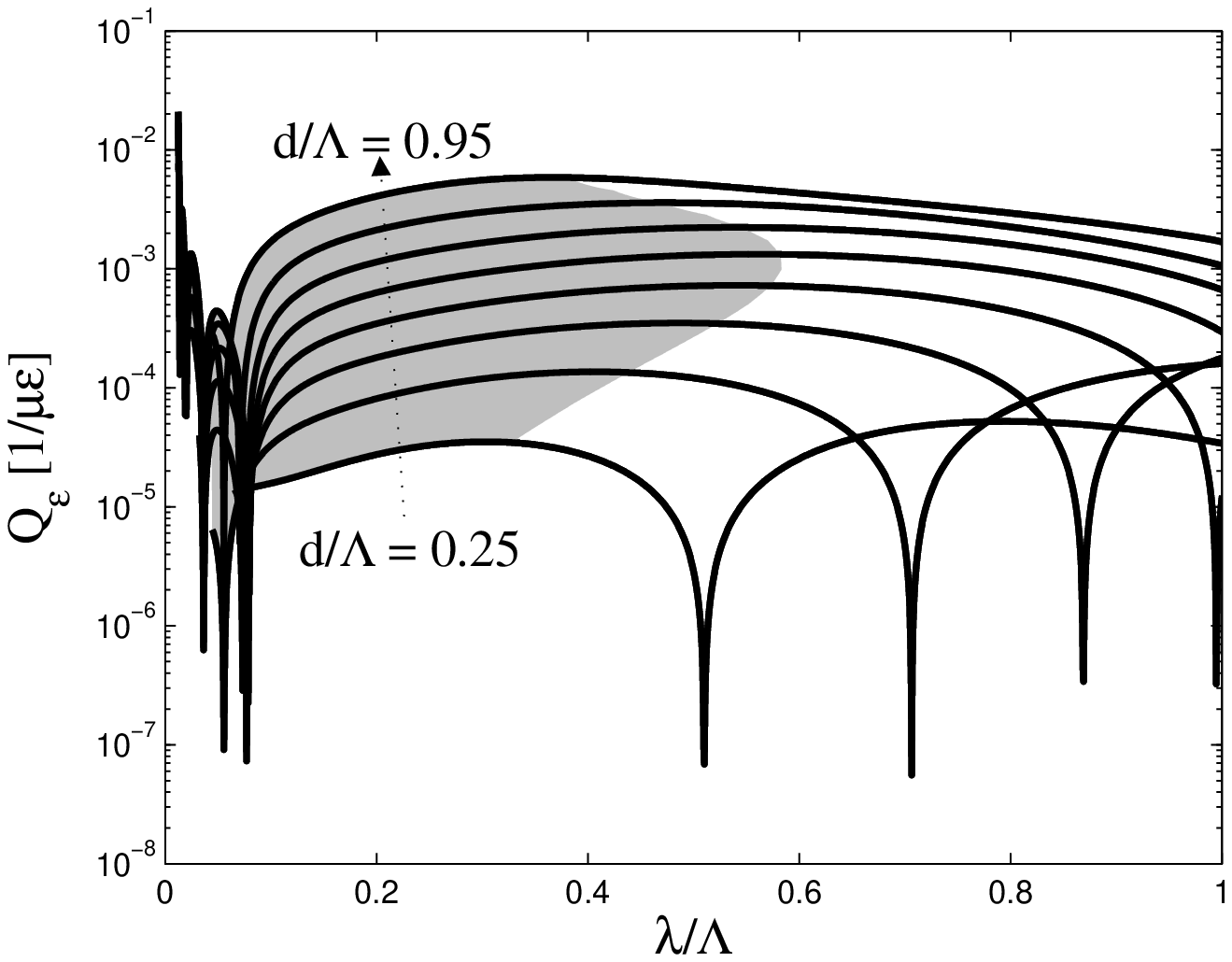}
\end{center}
\caption{Quality factor, $Q_{\varepsilon_s}$, for LPG strain
sensing. The lines indicate different values of the hole diameter
relative to the pitch: 0.25 to 0.95 in steps of 0.10. The gray area
indicates negative group index mismatch $\bar n _\textrm{g} < 0$.
The wavelength is 1 $\mu$m, the length of the gratings is $L$ = 30
mm, and $|\kappa L | = \pi/(2L)$}\label{fig:lpg_strain_Q}
\end{figure}
\begin{figure}[b!]
\begin{center}
\includegraphics[angle = 0, width =0.45\textwidth]{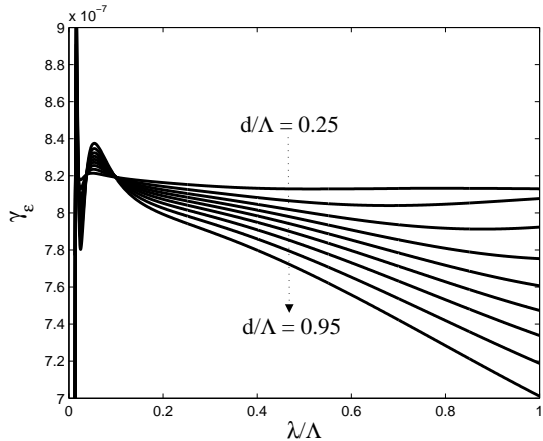}
\end{center}
\caption{Sensitivity ,$\gamma_{\varepsilon_s}$, for BG strain
sensing. The lines indicate different values of the hole diameter
relative to the pitch: 0.15 to 0.95 in steps of 0.10. The wavelength
is 1 $\mu$m, the length of the gratings is $L$ = 30 mm, and $|\kappa
L | = 4$}\label{fig:bg_strain_sens}
\end{figure}

Eqs.~(\ref{eq:bgstrain},\ref{eq:lpgstrain}) are inserted in the
expression for the sensitivity, Eq.~(\ref{eq:gamma}) and calculated
for the wavelength 1.0 $\mu$m. The results are shown in
Figs.~\ref{fig:bg_strain_sens} and \ref{fig:lpg_strain_sens},
respectively. The sensitivity for BGs  is very close to being
constant regardless of pitch,  hole diameters, and wavelength, as
seen in Fig.~\ref{fig:bg_strain_sens} and
Table~\ref{table:temp_lin}.  This can understood by that fact that
$n_\textrm{co}$ and $n_\textrm{g,co}$  are of similar magnitudes,
such that $n_\textrm{co} \gg \Lambda\frac{\partial
n_\textrm{co}}{\partial \Lambda}$. We can neglect the second term in
Eq.~\ref{eq:bgstrain}.
  We  then put
$\Delta\bar n_\textrm{f} / \varepsilon_s =
(1+{\eta_\varepsilon}/{n_b} )n_\textrm{g,co} \simeq 0.8 \times
n_\textrm{g,co}$, where $n_b\simeq 1.45$ is the refractive index of
silica. Inserting this in the expression for the sensitivity,
Eq.~(\ref{eq:gamma}), we realize that
$\gamma_{\varepsilon_s}/\mu\varepsilon_s \approx 0.8\times10^{\rm
\scriptsize -6}$ and $\Delta\lambda = 0.8\times10^{\rm \scriptsize
-6}\,\lambda\, \mu\varepsilon_s$. This result is in very good
agreement with the values for BGs displayed in table
\ref{table:strain_lin} and in Fig.~\ref{fig:bg_strain_sens}.

The quality factor, $Q_{\varepsilon_s}$, for LPGs with
$\lambda/\Lambda > 0.2$ tends to increase for increasing hole size,
while being almost independent of wavelength,  as seen in
Fig.~\ref{fig:lpg_strain_Q}. For these wavelength the first term in
Eq.~(\ref{eq:lpgstrain}), the elongation of the PCF, is dominating
over the strain-optic effect and the lateral strain.  For
$\lambda/\Lambda < 0.2$ the sensitivity is generally lower since
$\bar n_\textrm{f}$ decrease for decreasing wavelengths, but also
because the strain-optic effect and the lateral strain become
significant. The dips indicate a net cancelation of the terms in
Eq.~(\ref{eq:lpgstrain}).

 To validate our theory we compare with experiments. Dobb et al.
\cite{dobb2006} characterized LPGs in PCFs inscribed with an
electric arc. For a PCF with structure parameters $\Lambda \simeq
7.1 \mu$m, $d \simeq 4.0\mu$m they found a shift of resonance
wavelength with -2.5 pm/$\mu\varepsilon_s$  at the wavelength 1688
nm. Using the structure parameters in the theory yields -3.7
pm/$\mu\varepsilon_s$. For a PCF with structure parameters $\Lambda
\simeq 8 \mu$m, $d \simeq 3.7\mu$m, they found
-2.08/$\mu\varepsilon_s$ at 1403 nm wavelength and the theory gives
-3.2/$\mu\varepsilon_s$. Thus we have an agreement of the right
magnitude and right sign of the wavelength shift. Their PCF
structure values, $d/\Lambda \simeq 0.46, \lambda/\Lambda \simeq
0.18$, gives a $Q$ that is close to the dips in
Fig.~\ref{fig:lpg_strain_Q}. The dips corresponds to a cancelation
of the terms in Eq.~(\ref{eq:lpgstrain}) and close to the dips the
quality factor is more susceptible to errors between experimental
and numerical parameters.
 Martelli et al. used a PCF
with a erbium-doped UV sensitive core to inscribe BGs
\cite{martelli2005}. Their structure parameters are roughly $\Lambda
\simeq 10 \mu$m, $d \simeq 2\mu$m.
 At a wavelength of 1535 nm the wavelength shift is
1.2 pm/$\mu\varepsilon_s$. The theory gives 1.25
pm/$\mu\varepsilon_s$ for a pure silica PCF. A doped core does not
change the strain-optic coefficient significantly due to a doped
core \cite{park2002}.

For BGs  in a small core fiber with the structure parameters
$d/\Lambda \simeq 0.5$ and  $\Lambda \simeq 1.6 \mu$m
\cite{frazao2005} it is found experimentally a sensitivity of 1.25
pm/$\mu\varepsilon$ at the wavelength 1508 nm. The core was doped
with germanium to make the core UV-sensitive. Our theory gives 1.16
pm/$\mu\varepsilon$ in good agreement with the experimental value.

In conclusion we have found that the sensitivity for BGs is
approximately constant both for temperature and strain sensing, with
the coefficients $\gamma_T \simeq 0.62 \times 10.6/$K and
$\gamma_{\mu\varepsilon_s} \simeq 0.8\times 10^{\rm \scriptsize -6}$
for pure silica PCFs. The wavelength shift both for BGs and LPGs is
generally of the order of 5 pm/K, and the sensitivity and quality
factor are an order of magnitude large (in K) compared with strain
(in microstrain). The quality factor for LPGs for temperature and
strain sensing has roughly the same characteristics.


\section{Conclusion}
In conclusion we have shown that the performance of LPG sensors are
best characterized by the detectability, $Q$, rather than the
sensitivity, $\gamma$. We have calculated expressions for the
sensitivity, that incorporates the material-optic responses to the
external perturbation, as well as treated effects of having air
holes in the structure. Using the results of a fully-vectorial
finite element calculations, incorporating material dispersion, the
sensitivity and detectability have been calculated for a
range of PCF structure parameters, i.e. hole diameters and pitch
lengths.

We have proposed expressions for the sensitivity for internal
refractive index sensing, label-free biosensing, strain, and
temperature sensing. For strain and temperature we have included
both thermo-optic and strain-optic effects, as wells as the effects
of both longitudinal and lateral elongation. The formulas are of
general form and apply to all PCFs as well as standard optical
optical fibers.

For PCFs infiltrated with samples, the $Q$ for refractive index
sensing and biosensing may be improved by several orders of
magnitude by appropriate choice of the hole size and pitch. For
temperature and strain index sensing with BG sensors we have shown
that the sensitivity is almost constant for all PCFs. For LPGs an
accurate theoretical prediction of the sensitivity for temperature
sensing or strain sensing can be difficult since the sensitivity
strongly affected by the waveguide dispersion. However, our analysis
shows that the low temperature sensitivity of PCF-gratings,  $\sim
$5 pm/K, is generally valid for all PCFs.

L. Rindorf's e-mail address is lhr@com.dtu.dk.

\section{Appendix: First order perturbation theory}
The linear change in propagation constant by a small perturbation in
the dielectric function, $\Delta\varepsilon$, is \cite{snyder}
\begin{eqnarray}
\Delta\beta_m = \frac {\omega} {2}  \frac {\partial \beta} {\partial
\omega} \frac{\int_\Omega \hspace{-0.1cm}d\rrr_\bot
\Delta\varepsilon(\rrr_\bot)|\EEE_m(\rrr_\bot)|^2}{\int_\Omega
\hspace{-0.1cm}d\rrr_\bot
\varepsilon(\rrr_\bot)|\EEE_m(\rrr_\bot)|^2}.
\end{eqnarray}
$\Omega$ denotes the fiber cross section. By setting the norm of the
orthogonal solutions to the Helmholtz equation (\ref{eq:eigenvalue})
to unity,
\begin{eqnarray}
\int_\Omega \ud\rrr_\bot
\EEE^\dagger_m(\rrr_\bot)\varepsilon(\rrr_\bot)\EEE_n(\rrr_\bot)=
\delta_{mn},
\end{eqnarray}
where $\dagger$ denotes Hermitian conjugate (the complex conjugate
transposed), we get
\begin{eqnarray}
\Delta n_{textrm{eff},m} = \frac{n_{\textrm{g},m} } 2 \int_\Omega
\hspace{-0.1cm}\ud\rrr_\bot
\Delta\varepsilon(\rrr_\bot)|\EEE_m(\rrr_\bot)|^2
\end{eqnarray}
with the group index, $n_{\textrm{g},m} \equiv {\rm c} \frac
{\partial \beta} {\partial \omega}$. We define the field energy
intensity in the air holes of the mode $m$  by
\begin{eqnarray} f_{u,m} \equiv\frac{
\int_{\Omega_{\rm h}}d\rrr_\bot \DDD_m^\dagger(\rrr_\bot ) \cdot
\EEE_m(\rrr_\bot )}{\int_{\Omega}d\rrr_\bot \DDD_m^\dagger
(\rrr_\bot )\cdot \EEE_m(\rrr_\bot )}.
\end{eqnarray}
$\Omega_{\rm h}$ is the part of the fiber cross section covering
holes, and $\DDD = \varepsilon \EEE$. The linear change in effective
index of mode $m$ caused by a change of the refractive index of the
holes is then
\begin{eqnarray}
\frac{\partial n_{\textrm{eff},m}}{\partial n_{\rm h}} &=&
\frac{n_{\textrm{g},m}}{n_{\rm h}}f_{u,m}
\label{eq:deltaneff_refrac_app_hole}
\end{eqnarray}
where $n_{\rm h}$ is the refractive index of the holes. The change
is proportional to $f_{u,m}$ The corresponding linear change in
effective index caused by a change of the refractive index of the
base material is
\begin{eqnarray}
\frac{\partial n_{\textrm{eff},m}}{\partial n_{\rm b}} &=&
\frac{n_{\textrm{g},m}}{n_{\rm b}}(1-f_{u,m})
\label{eq:deltaneff_refrac_app_base}
\end{eqnarray}
where $n_{\rm b}$ is the refractive index of the base material. It
is seen that the change is proportional to the field intensity in
the base material, $1-f_u$.


\begin{thebibliography}{10}
\newcommand{\enquote}[1]{``#1''}

\bibitem{lee2003}
B.~Lee, \enquote{Review of the present status of optical fiber
sensors,}
  Optical Fiber Technology \textbf{9}, 57--79 (2003).

\bibitem{birks1997}
T.~A. Birks, J.~C. Knight, and P.~S.~J. Russell, \enquote{Endlessly
single mode
  photonic crystal fibre,} Opt. Lett. \textbf{22}, 961--963 (1997).

\bibitem{eij2001}
M.~A. van Eijkelenborg, M.~C.~J. Large, A.~Argyros, J.~Zagari,
S.~Manos,
  N.~Issa, I.~Bassett, S.~Fleming, R.~C. McPhedran, C.~M. de~Sterke, and
  N.~A.~P. Nicorovici, \enquote{Microstructured polymer optical fibre,} Opt.
  Express \textbf{9}, 319 -- 327 (2001).

\bibitem{humbert2003}
G.~Humbert, A.~Malki, S.~Fevrier, P.~Roy, and D.~Pagnoux,
\enquote{Electric
  arc-induced long-period gratings in ge-free air-silica microstructure
  fibres,} Electron. Lett. \textbf{4}, 349--350 (2003).

\bibitem{kakarantzas2002}
G.~Kakarantzas, T.~A. Birks, and P.~S.~J. Russell,
\enquote{Strucural
  long-period gratings in photonic crystal fibers,} Opt. Lett. \textbf{27},
  1013--1015 (2002).

\bibitem{nielsen2003}
M.~D. Nielsen, G.~Vienne, J.~R. Folkenberg, and A.~Bjarklev,
  \enquote{Investigation of micro deformation induced attenuation spectra in a
  photonic crystal fiber,} Opt. Lett. \textbf{28}, 236--238 (2003).

\bibitem{eggleton:1999}
B.~J. Eggleton, P.~S. Westbrook, R.~S. Windeler, S.~Sp\"alter, and
T.~A.
  Strasser, \enquote{Grating resonances in air-silica microstructured optical
  fibers,} Opt. Lett. \textbf{24}, 1460 -- 1462 (1999).

\bibitem{peng1999}
G.~D. Peng, Z.~Xiong, and P.~L. Chu, \enquote{Photosensitivity and
gratings in
  dye-doped polymer optical fibers,} Opt. Fiber. Technol. \textbf{5}, 242--251
  (1999).

\bibitem{li2005}
Z.~Li, Y.~Tam, L.~Xu, and Q.~Zhang, \enquote{Fabrication of
long-period
  gratings in poly(methyl methacrylate-co-methyl vinyl ketone-cobenzyl
  methacrylate)-core polymer optical fiber by use of a mercury lamp,} Opt.
  Lett. \textbf{30}, 1117--1119 (2005).

\bibitem{dobb2005}
H.~Dobb, D.~Webb, K.~Kalli, A.~Argyros, M.~Large, and M.~van
Eijkelenborg,
  \enquote{Continuous wave ultraviolet light-induced fiber bragg gratings in
  few- and single-mode microstructured polymer optical fibers,} Opt. Lett.
  \textbf{30}, 3296--3298 (2005).

\bibitem{hiscocks2006}
M.~P. Hiscocks, M.~A. van Eijkelenborg, A.~Argysor, and M.~C.~J.
Large,
  \enquote{Stable imprinting of long-period gratings in microstructured polymer
  optical fibre,} Opt. Express \textbf{14}, 4644--4649 (2006).

\bibitem{frazao2005}
O.~Frazao, J.~P. Carvalho, L.~A. Ferreira, F.~M. Araujo, and J.~L.
Santos,
  \enquote{Discrimination of strain and temperature using bragg gratings in
  microstructured and standard optical fibres,} Meas. Sci. Technol.
  \textbf{16}, 2109--2113 (2005).

\bibitem{dobb2004}
H.~Dobb, K.~Kalli, and D.~Webb, \enquote{Temperature-insensitive
long period
  grating sensors in photonic crystal fibre,} Electron. Lett. \textbf{11},
  657--658 (2004).

\bibitem{sorensen2006}
H.~R. Sorensen, J.~Canning, J.~Laegsgaard, and K.~Hansen,
\enquote{Control of
  the wavelength dependent thermo-optic coefficients in structured fibres,}
  Opt. Express \textbf{14}, 6428--6433 (2006).

\bibitem{delisa2000}
M.~P. DeLisa, Z.~Zhang, M.~Shiloach, S.~Pilevar, C.~C. Davis, J.~S.
Sirkis, and
  W.~E. Bentley, \enquote{Evanescent wave long-period fiber bragg grating as an
  immobilized antibody biosensor,} Analytical Chemistry \textbf{72}, 2895--2900
  (2000).

\bibitem{jensen2004a}
J.~B. Jensen, L.~H. Pedersen, P.~E. Hoiby, L.~B. Nielsen, T.~P.
Hansen, J.~R.
  Folkenberg, J.~Riishede, D.~Noordegraaf, K.~Nielsen, A.~Carlsen, and
  A.~Bjarklev, \enquote{Photonic crystal fiber based evanescent-wave sensor for
  detection of biomolecules in aqueous solutions,} Opt. Lett. \textbf{29},
  1974--1976 (2004).

\bibitem{jensen2005}
J.~B. Jensen, P.~E. Hoiby, G.~Emiliyanov, O.~Bang, L.~H. Pedersen,
and
  A.~Bjarklev, \enquote{Selective detection of antibodies in microstructured
  polymer optical fibers,} Opt. Express \textbf{13}, 5883--5889 (2005).

\bibitem{gem2007}
G.~Emiliyanov, J.~B. Jensen, O.~Bang, P.~E. Hoiby, L.~H. Pedersen,
E.~M. Kjaer,
  and L.~Lindvold, \enquote{Localized biosensing with topas microstructured
  polymer optical fiber,} Opt. Lett. \textbf{32}, 460--462 (2007). Opt. Lett. \textbf{32}, 1059
  (2007).

\bibitem{kerbage2003}
C.~Kerbage and B.~Eggleton, \enquote{Tunable microfluidic optical
fiber
  gratings,}  \textbf{82}, 1338--1340 (2003).

\bibitem{rindorf2006_7}
L.~Rindorf, J.~B. Jensen, M.~Dufva, L.~H. Pedersen, P.~E. Hoiby, and
O.~Bang,
  \enquote{Biochemical sensing using photonic crystal fiber long-period
  gratings,} Opt. Express \textbf{14}, 8824--8831 (2006).

\bibitem{phanhuy2006}
M.~C.~P. Huy, G.~Laffont, Y.~Frignac, V.~Dewynter-Marty,
P.~Ferdinand, P.~Roy,
  J.-M. Blondy, D.~Pagnoux, W.~Blanc, and B.~Dussardier, \enquote{Fibre bragg
  grating photowriting in microstructured optical fibres for refractive index
  measurement,} Meas. Sci. Technol. \textbf{17}, 992--997 (2006).

\bibitem{rindorf2006_6}
L.~Rindorf, P.~E. Hoiby, J.~B. Jensen, L.~H. Pedersen, O.~Bang, and
O.~Geschke,
  \enquote{Towards biochips using microstructured optical fiber sensors,}
  Analytical and Bioanalytical Chemistry  (2006).

\bibitem{acharya}
B.~R. Acharya, T.~Krupenkin, S.~Ramachandran, Z.~Wang, C.~C. Huang,
and J.~A.
  Rogers, \enquote{Tunable optical fiber devices based on broadband long-period
  gratings and pumped microfluidics,}  \textbf{83}, 4912--4914 (2003).

\bibitem{ramachandran2005}
S.~Ramachandran, \enquote{Dispersion-tailored few-mode fibers: A
versatile
  platform for in-fiber photonic devices,} J. Lightwave Technol. \textbf{23},
  3426--3443 (2005).

\bibitem{kuhlmey2002}
B.~T. Kuhlmey, R.~C. McPhedran, and C.~M. {de Sterke},
\enquote{Modal cutoff in
  microstructured optical fibers,} Opt. Lett. \textbf{27}, 1684--1686 (2002).

\bibitem{mortensen2003c}
N.~A. Mortensen, J.~R. Folkenberg, M.~D. Nielsen, and K.~P. Hansen,
  \enquote{Modal cutoff and the $v$ parameter in photonic crystal fibers,} Opt.
  Lett. \textbf{28}, 1879--1881 (2003).

\bibitem{yariv_book}
A.~Yariv and P.~Yeh, \emph{Photonics} (Oxford University Press,
Oxford, 2007).

\bibitem{shu2002}
X.~W. Shu, L.~Zhang, and I.~Bennion, \enquote{Sensitivity
characteristics of
  long-period fiber gratings,} J. Lightwave Technol. \textbf{20}, 255--266
  (2002).

\bibitem{comsol}
Http://www.comsol.com.

\bibitem{water}
Http://www.iapws.org/relguide/rindex.pdf.

\bibitem{daxhelet2003}
X.~Daxhelet and M.~Kulishov, \enquote{Theory and practice of
long-period
  gratings: when a loss becomes a gain,} Opt. Lett. \textbf{28}, 686--688
  (2003).

\bibitem{dobb2006}
H.~Dobb, K.~Kalli, and D.~Webb, \enquote{Measured sensitivity of
arc-induced
  long-period grating sensors in photonic crystal fibre,} Opt. Com.
  \textbf{260}, 184--191 (2006).

\bibitem{martelli2005}
C.~Martelli, J.~Canning, N.~Groothoff, and K.~Lyytikainen,
\enquote{Strain and
  temperature characterization of photonic crystal fiber bragg gratings,} Opt.
  Lett. \textbf{30}, 1785--1787 (2005).

\bibitem{park2002}
Y.~Park, T.-J. Ahn, Y.~H. Kim, W.-T. Han, U.-C. Paek, and D.~Y. Kim,
  \enquote{Measurement method for profiling the residual stress and the
  strain-optic coefficient of an optical fiber,} Appl. Opt. \textbf{41}, 21--26
  (2002).

\bibitem{snyder}
A.~W. Snyder and J.~D. Love, \emph{Optical Waveguide Theory}
(Chapman \& Hall,
  New York, 1983).

\end{thebibliography}

\end{document}